\documentclass[pre,twocolumn,aps,floatfix,superscriptaddress]{revtex4}

\usepackage[english]{babel}
\usepackage{amssymb,amsfonts,amsmath}
\usepackage[pdftex]{graphicx}
\usepackage{latexsym}
\usepackage{multirow}
\usepackage{color}
\usepackage{ltxtable}

\usepackage{amsmath}
\usepackage{hyperref}

\begin{document}


\title{Complexity, Selectivity and Asymmetry  in the Conformation of the Power Phenomenon. Analysis of Chilean Society}

\author{Juan Pablo C\'ardenas}
\affiliation{Inria-Chile. Apoquindo 2827, Piso 12, Las Condes. Santiago, Chile.}
\affiliation{Centro de Investigaci\'on en Complejidad Social. Facultad de Gobierno. Universidad del Desarrollo. Av. Plaza 680, San Carlos de Apoquindo, Santiago, Chile.}

\author{Gerardo Vidal}
\affiliation{Centro de Estudios Estrat\'egicos de la Armada de Chile. Tom\'as Ramos, 12, Valpara\'iso, Chile.}
\author{Gast\'on Olivares}
\affiliation{Instituto de Sistemas Complejos de Valpara\'iso, Subida Artiller\'ia 470, Valpara\'iso,  Chile.}

\begin{abstract}

In this work we analyzed the relationships between powerful politicians and businessmen of Chile in order   to study the phenomenon of social power.  We developed our study according to Complex Network Theory but also using traditional sociological theories of Power and Elites.   Our   analyses suggest that the  studied network displays common properties of Complex Networks, such  as  scaling  in  connectivity distribution,  properties of \textit{small-world}  networks, and modular structure,  among others.  We  also  observed that  social  power  (a  proposed metric is  presented in  this  work)   is  also  distributed inhomogeneously. However, the  most  interesting observation is that this  inhomogeneous power  and  connectivity distribution, among other observed  properties, may  be the  result of a dynamic and  unregulated process  of network growth in which  powerful people  tend to  link  to  similar others.   The  compatibility between people, increasingly selective as  the network grows, could  generate the presence of extremely powerful people, but also a constant inequality of power  where  the  difference between the  most  powerful is the  same  as among the  least powerful. Our results are also in accordance with  sociological theories.   

\vspace{0.5 cm}

\textbf{keywords}: Complex Networks, Social Networks, Model, Theory of Elites, Theory of Social Power.

\end{abstract}

\maketitle

\section{Introduction}

Perhaps nothing in human history has been more seductive than power and its relationships. It seems very likely that its manifestation is parallel to the act of abandon the (primal) isolation and to form a society in order to achieve, in common, its protection and survival. The sociological literature is replete with the power phenomenon, and it has been explained in many ways \cite{Molina}; these explanations are based  not  so much on its explicit  expression (imposition  of will over another, or even against  its resistance \cite{Weber,Lenski}), but rather on its relational character (\textit{i.e.}, there's  always someone who commands  and  another  who obeys) and very particularly on the factors that make such interaction possible (both for those  who rule and those who obey). 

The social sciences, and political sociology in particular, have paid attention to the study of power, particularly on ``what to do'' once achieved, even, subordinating what should be done once it has been achieved. This comes from a long tradition first expressed in the second century B.C. by  Polybius \cite{Touchard} who prioritizes a pragmatic sense of politics subordinating moral purpose. Niccol\`o Machiavelli, republished this  position of decoupling  politics  from  ethics thirteen centuries  later  \cite{Touchard}. Since then,  it has been  power, and  not  the  common  good, which has  preferentially concerned the  literature of the social sciences.  More recently, however, and perhaps  in an attempt to explain  (and modify) society itself, other frameworks, such as the Theory of Organizations and General Systems Theory  \cite{Bertalanffy} have made contributions to the phenomenon from a systemic  point of view\footnote{These theories have explored the phenomenon of power, giving explanations of the  process of the gains or losses thereof and  the subsequent effects  on the  structure and  efficiency  of the  systems.}. One example of this kind of approach is the work done by Mark Lombardi, an American artist and historian who studied social power, or more clearly its ``uses and abuses'' \cite{Bartolo}. He will be remembered not only due to his art but  for his bravery  and  for his work's non-conventional way of looking at the  phenomenon. In Lombardi's view, the power of people cannot be understood by looking at individuals; rather, it is necessary  to look at  the system. In fact, his work was immortalized in his graph drawings called Narrative Structures, where people and organizations are linked by power relations. Today, this kind of approach is common because social scientists recognize the limitations of reductionism in addressing social problems. Notwithstanding these modern procedures, it was in 18$^{th}$ century that E. Durkheim \cite{Ritzera,Deploige}  first opened the door to understanding sociological phenomena from this perspective. He talked about a concept called ``dynamic density'' in order to explain the transition of an organic society to a mechanical one. This change is produced by a significant increase in the number of actors of the system and simultaneously, the number of interactions between them. Nowadays, this is known as a fundamental characteristic of Complex Systems and the evidence that this kind of social complex structures display collective behaviors that are not described in the actors, makes it necessary its study from an holistic perspective, since it is the web of relationships who ``hides'' the information about the system's properties \cite{Buchanan, Ball, Taleb}. 
This coincides with the ideas of Michel Foucault \cite{Foucault} on social power: ``Power must be analyzed as something which circulates, or rather as something that operates in chain. It is never localized here or there, it is never in the hands of someone, it is not an attribute such as wealth or goods. The power functions... and is exercised through networks...''.

In 2012 a Chilean website, called Poderopedia, was founded on the same spirit of Lombardi`s work. Supported by a 16-month News Challenge Grant provided by the Knight Foundation\footnote{\url{http://www.knightfoundation.org/grants/20110953/}}, and later with the help of Start-Up Chile\footnote{\url{http://www.startupchile.org/congrats-ya?ll/}} and new funds provided by the Knight Fellows Program of the International Center for Journalists (ICFJ)\footnote{\url{http://www.icfj.org/}}, Poderopedia has developed a huge public data base with (public) information about relations between powerful businessmen, politicians and organizations/companies in Chile with the aim of providing public transparency as a project embedded in what appears to be a global need. In fact, Poderopedia's spirit seems to capture a concern of much of the world's population and recently the website is already operating in Colombia and Venezuela. However, the data collected by Poderopedia are also interesting from a scientific point of view, not only because of the amount (and dynamics) of the data, but also due to its richness, which might  contains (hide) information about the power phenomenon. Precisely that motivated this work. Using the data from Poderopedia, we studied the social network of Chilean powerful people; this was not  with the aim of detecting  or reporting the powerful, but rather to map social power in Chile, characterizing it according to the links that define the relation among people, and trying to explain the origin of its structure.

Considering the huge number of people and organizations/companies intricately connected, it was necessary to approach the problem using tools from Complex Systems Theory while also relying on the sociological theory of Power and Elites. This effort was based on the evidence that certain features of complex systems are ideal in understanding and explaining certain social processes more comprehensively. 
What is surprising is that using a novel approach to the social power phenomenon, we found a principle of causal interpretation derived from Complex Network theory. This principle was contrasted with the major political underpinnings, discovering, in this kind of ``assisted reproduction'' among disciplines, surprising consequences. Our work puts in evidence that the power phenomenon differs little from other collective behaviors of completely different nature. In fact, the complexity of the power system seems to converges to a universal architecture that depends on the system interactions rather than psycho-social or even moral factors affecting the political and business world.

The  work is structured as follows: in the  next  section, the  methodology used by Poderopedia to complete  its relational data is presented and,  in the  same section, we also show how we process  the  data  provided  in order  to obtain  graphs  of entities which map  the  structures of the  social power network  of Chile;  section  3 introduces the  social network  model  that we propose  in order  to  explain  the  structure observed  in the  real network from section 2; in Section 4, we present a sociological interpretation of the  results  obtained in the  previous  sections; and, in the  last  section, we discuss the major implications  of the obtained results  and present the conclusions of our work.

\section{Chilean Social Power Network: Construction and Analysis}\label{chile_net}

Poderopedia is a collaborative data journalism platform (\url{poderopedia.org}) that maps the ``who's who in business and politics in Chile''. It is developed  by journalists, programmers,  designers, and independent collaborators who investigate, extract, select, and validate  information from public sources, such as media, government databases, business databases, and  websites.   This information is subjected to careful  source verification processes and the most relevant aspects  of public interest are stored and published.

Poderopedia was designed to be used as a tool in the day-to-day of journalists, media,  general public, professionals,  organizations, and companies  who need to find information about  the relations between powerful people and  organizations in Chile.   It  also seeks to promote   citizen  participation by giving  users  the  option  to  register  on  the website to provide  all kinds of data.

The data collected are stored according to the ontology called PoderVocabulary (\url{https://github.com/poderopedia/podervocabulary}), specially designed for this purpose. PoderVocabulary is based on OWL (Web Ontology Language), Friend-of-a-friend (FOAF) and BIO, an extension of FOAF, focused on biographical information.  Data  stored  can  be accessed  by users  who want to search  for specific entities' properties  as well as their  relations.  Users registered  in Poderopedia can also suggest new entity relationships.

Entities in the Poderopedia ontology can be of two types:  Persons and Organizations/Companies; each is characterized by a set of attributes that describe their properties. Furthermore, relations  between entities  are also of different kinds: Person-Person, Person-Organization, and Organization-Organization. A sub set of connection types are defined inside of each link category (Table  \ref{tab1}).

\begin{table}[h]
 \caption{Types  of relationships used by Poderopedia to link entities.}
\begin{tiny}
 \begin{tabularx}{250pt}{XX}
\hline \\ 
Relation 	category			& Relation type \\
\hline \\ 
Person-Person			& family, friend, close, known, classmate\\
Person-Organization 	& position in company, participation in company,\\
& position in NGO, participation in NGO, \\
&participation in economic group, \\
&position in International organism, \\
&public office, study, religious group, \\
&support groups for political campaign, \\
&social movement, private club, \\
&political party, think tanks or study center \\
Organization-Organization 	& commercial, dependence, donations, property, grants\\

\end{tabularx}
\label{tab1}
\end{tiny}
\end{table}

We worked  with  data  stored  in Poderopedia servers  between June 2012 and March 11$^{th}$, 2014. The reason is that day the change of government took place in Chile. Michelle Bachelet began her second period as President and Sebasti\'an Pi\~nera finished his one. Any data collected after  that date  might have created noise in our analysis due to the  entry  of many  new entities and retalionships  (associated with the  new government) to the system. 

Our analysis was initially focused on person to person relations. For this reason, we projected the original network onto a unimodal structure. Since the original network of Poderopedia is not a real bipartite graph\footnote{\textit{theyrule.net} is a website similar to  Poderopedia, but offers  real  bipartite graphs, where  people  are  not  linked  among themselves but through organizations/companies.}, the projection made is not one typically used in the case of bipartite networks; that is, there  are links between entities of the same type. Thus, if two persons  are  connected  by  an  organization/company  in  the  original network,  they would be connected  in the projected  network.  However, if two persons are connected by a path composed of two adjacent organizations/companies, they are not connected (through this path) in our projected network.

\begin{figure}[htp]
\begin{center}
  \includegraphics[width=3in]{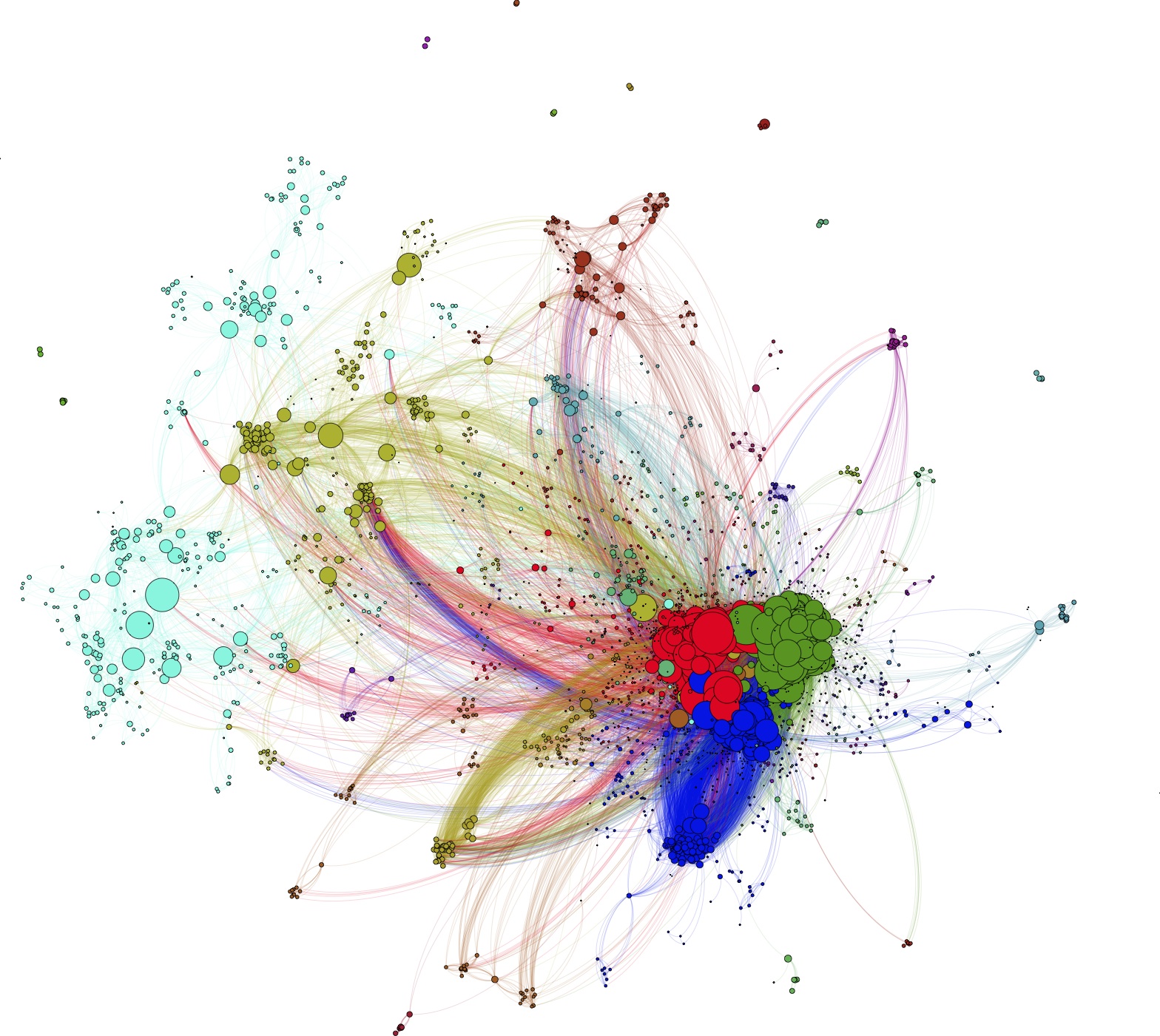}\\
  \caption{Network of Chilean Power. The network represents the projected network of 3140 people and 17353 edges. The layout, obtained using the OpenOrd algorithm (\url{https://github.com/gephi/gephi-plugins/tree/openord-layout}), shows the powerful people (see definition below)  bigger than the rest and colored by community partition using the Blondel \textit{et al.} algorithm \cite{Blon}.}
  \label{fig1}
  \end{center}
\end{figure}

Figure \ref{fig1} shows the projected people network. This network is composed of $N$=3140 people and $E$=17353 edges between them. The figure also shows nodes and links colored according to the community partition detected using the algorithm proposed by Blondel \textit{et al.} \cite{Blon}. Although we treat the networks as undirected, the color of links represents the community color of the source node according to the relational data used.

We were interested in the projected people network because we needed to know what was the most powerful people of the network, and then analyzed their relationships as cases of study. Although  the  social power in this  network  could be rudimentarily defined as the number  of incoming and outgoing  links, we wanted  to find a more sophisticated metric.  For this  reason  we implemented a metric of social power, $P$, that considers the quality of links of the person (Page Rank \cite{PR}), the cohesiveness of the immediate neighborhood (clustering coefficient, $c_i$), the capacity of the person to connect people (betweenness coefficient, $b_i$) and the proportion of person's links that represent family, business, religious or political relations. The sizes of the nodes in the network of Fig. \ref{fig1} represent corresponding $P$ values. 

\begin{figure}[htp]
\begin{center}
 \includegraphics[width=3in]{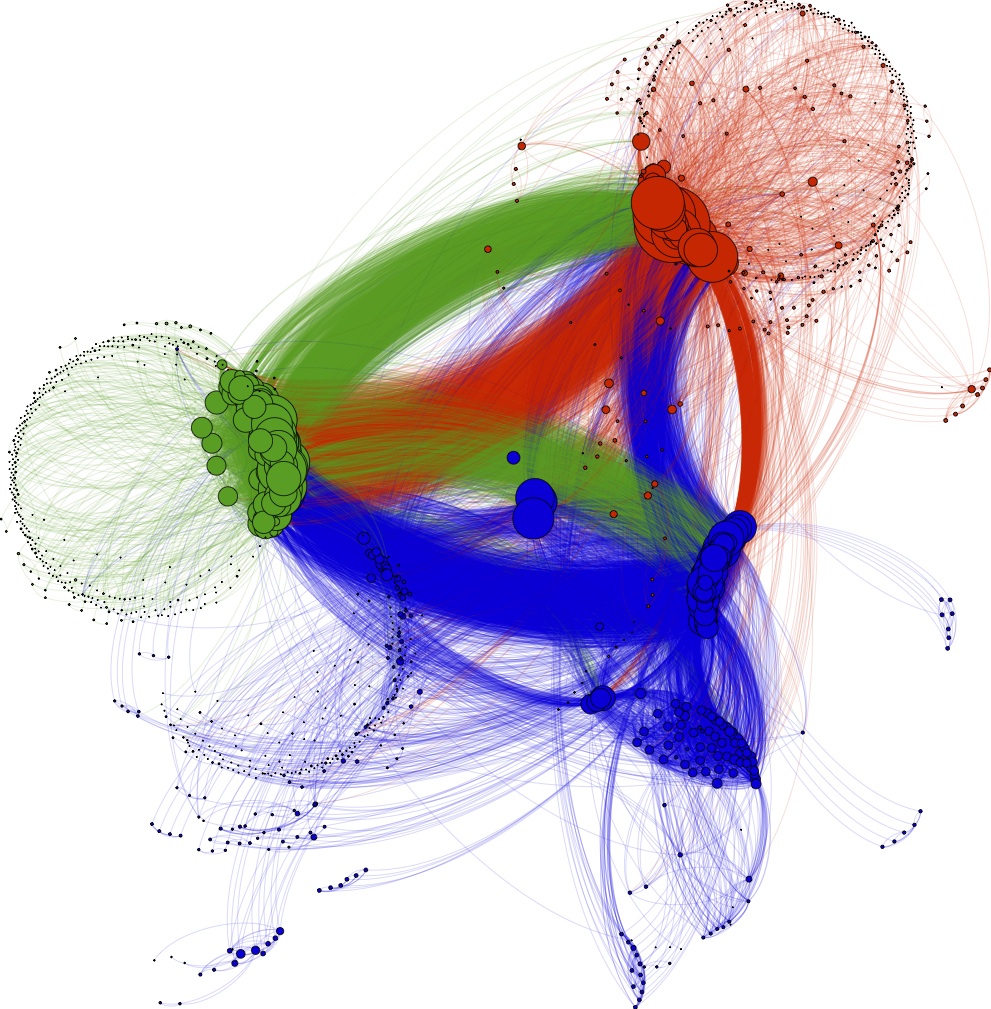}\\
\caption{The three major communities of the network of Chilean power. Color and node size as in Fig. \ref{fig1}.}
\label{fig2}
\end{center}
\end{figure}

The community analysis shows that there are three communities bigger than the rest, comprising 51.6\% of the nodes and 74.8\% of the links (red, blue and green communities in Fig. \ref{fig1}).  A detail of the connections  within this densely connected structure is shown in Figure \ref{fig2}. These  three  communities contain the set of people and relationships used for analysis in the following sections. Before that, however, let us now delve into some topological particularities of this network which are similar to those observed in other complex networks.

A wide range of real complex systems have been studied  from the  perspective  of Complex  Network  Theory.  This theory provides a framework for describing  interactions  in a system  from a purely topological  point  of view and allows us to abstract away dynamic processes that use such a structure as a substrate \cite{New}. A common and non-trivial topology was observed in most of the systems studied \cite{Newb,Dor}. In particular, the topology found in these complex networks was characterized by the presence of scaling in the node connectivity distribution.  This phenomenon denotes high inhomogeneity in connectivity, which is where the term \textit{scale-free} network comes from \cite{BA}; this is unlike the connectivity observed in networks where connections  are randomly  assigned.  Such inhomogeneity stems from the non-negligible presence  of densely connected  nodes  (\textit{hubs})  in these  networks. Besides inhomogeneity,  complex  topologies  share characteristic properties  of \textit{small-world}  networks \cite{WS}, \textit{i.e.}, they are associated  with low distances between randomly  chosen pairs of nodes and having a high clustering coefficient.  The \textit{scale-free} character and properties of \textit{small-world} networks seem to be a finger print of so-called complex networks \cite{Newb,Sole}.

\begin{figure}[htp]
\begin{center}
  \includegraphics[width=3in]{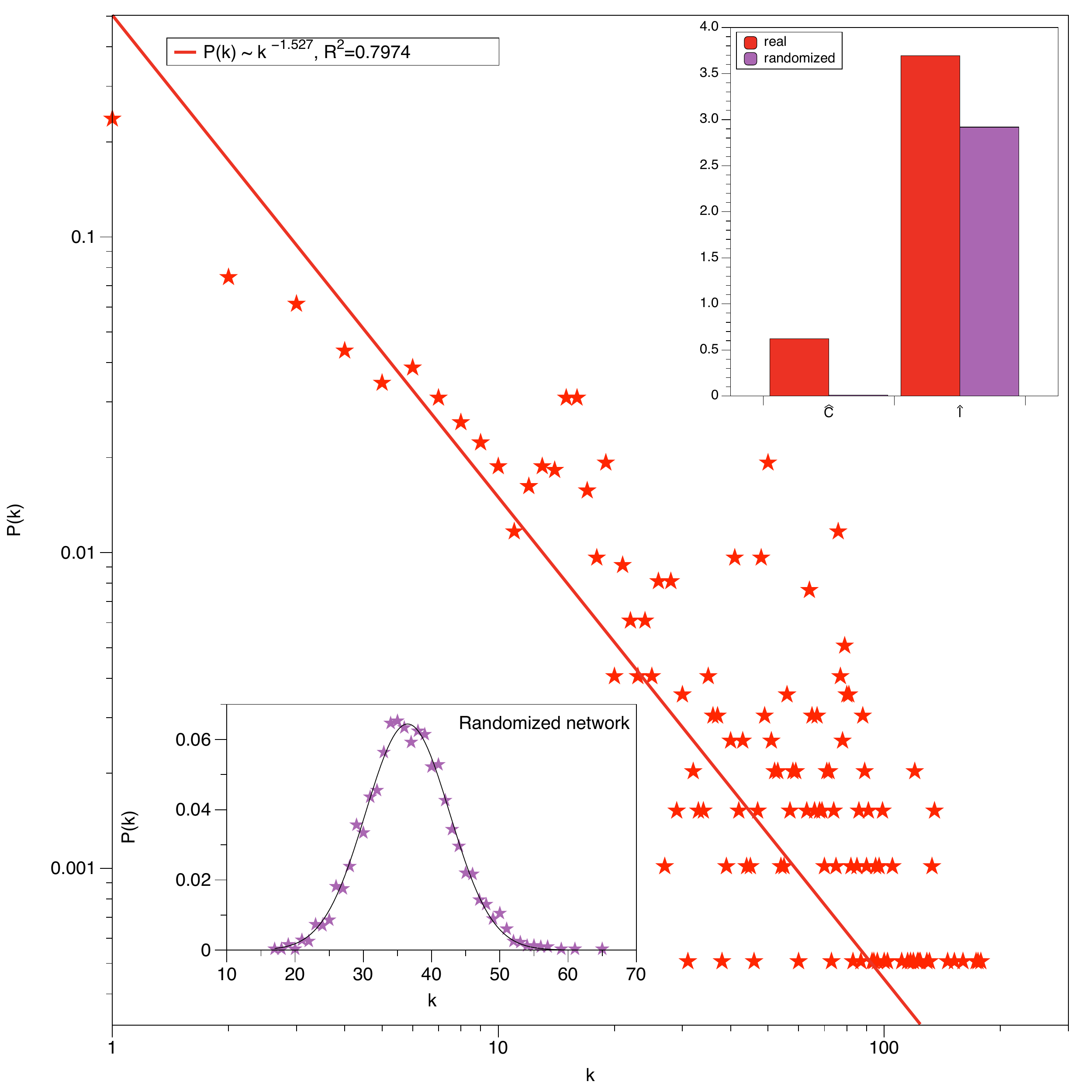}
   \caption{Node degree distribution, $P(k)$, for the real network and for its randomized version (bottom inset). Upper inset shows the mean clustering $\hat{C}$ and average path length $\hat{l}$ of the real network (red) and its randomized version (purple).}
  \label{fig3}
  \end{center}
\end{figure}

Using the Python \textit{powerlaw} package  \cite{Python}, we determined the function that best fits the connectivity distribution of our people network. The results show that the power law function is the best candidate, specifically, a truncated power law\footnote{Due  to  the  fact  that the  system has  a finite  size and  that there is a limit  of people  to  connect to,  the  powerful cannot become more powerful  forever. Thus it is probable that there is  a gradual upper bounding effect  on the  scaling  of the  power  law.   An  exponentially truncated power  law  is better suited to reflect  this  bounding \cite{Python}. Moreover, according to \cite{Bartolo}, this could be a product of ``prohibited relationships'', as in the case of mutualistic  networks in ecology \cite{Bascompte2}.} in comparison with normal, log-normal or exponential functions. This means that the people network displays the same \textit{scale-free} character that many  other  complex systems share, wherein  the  probability of choosing  a  person  with  $k$  connections  scales  (negatively) with the degree $k$. As can be seen in Fig. \ref{fig3}, close to 50\% of the nodes are poorly connected ($k\leq3$) but a few have many connections. This scenario, for a network  of the same size and same number  of links but  with random connections, is totally different (Fig. \ref{fig3} (bottom inside plot)). In the ``random'' scenario (not a real classical random graph \cite{ER} because the existing edges were randomly redirected), more than 80\% of the people have a connectivity close to the mean, between 30 and 40 links, and connectivities far from this magnitude are practically impossible (the probability decays following an exponential function). This  means  that the probability of finding a person  with  a connectivity higher (or less) than  the  mean  decays quickly  in comparison  with  the  scaling  observed  in the  original  network. In the real scenario the ``speed'' of probability decrease remain constant, that is, the inequality between densely connected  people is comparable to the one observed  between  the poorly connected  people.

Furthermore, we found that a short distance  between randomly  chosen pairs  of persons  coexists  with  a high transitivity of connectivity, as in \textit{small-world} networks \cite{WS}. The upper  inset of Figure \ref{fig3} shows how transitivity is lost in a random  scenario: the high transitivity of connection in the  social network ($\hat{C}$=0.62) practically goes to 0 when links are randomly distributed between  people.  By contrast, the average  path  length, $\hat{l}$, remains  practically  unaltered, suggesting that the  well known  \textit{small-world effect}, associated  with a short  average  path  length,  is not  only a property of complex graphs, but  also of those with random  connections \cite{ER}.

\section{Modeling the Social Power Network} \label{model}

Since the  social  network  of powerful Chileans shows properties reminiscent of complex  systems,  it follows that one should want to know the reason,  or at least the mechanism.  In this respect, we compared different dynamic  complex networks  models  that have been proposed in recent years \cite{Dor} .  One of them, the well-known Preferential Attachment model, or Barab\'asi-Albert (BA) model   \cite{BA}, have been the most popular. The model considers a social-like rule, where ``the rich get richer'', as the rule that governs the growth of the network. In this model,  a new node added to the system is linked preferentially to the most connected  nodes.  The mechanism  appears  to apply in the same way for the power network.  Indeed, this is intuitively how we as  non-powerful people think: power attracts more power.

However, this study has a different hypothesis.  According  to studies  concerning Elites and  Power  in society, ``a man (belonging to the Elite) is shaped by his relations with others like him...'' \cite{Mills}, ``(Elites) develop a common language and generate meeting spaces'' \cite{chilepoder} and  ``not enough that each elite has its own communication mechanisms, but it needed to be building a common image of society'' \cite{chilepoder}. In summary, it appears that the relationships between powerful people necessarily depends  on their  sharing  certain  characteristics. In fact, G. Tarde \cite{Nocera} talks about the \textit{phenomenon of imitation}, common among powerful people. Considering this, the Compatibility Attachment Model (CAM) \cite{CAM} seems to be a good candidate. 

In CAM, a (complex) network arises from \textit{compatibility}, a simple  local mechanism so called due to the fact that the relationship of two system entities  is a product of the compatibility between their characters. For the purposes of this study, compatibility would be this common language,  view of society, etc., common among  powerful persons.

In this model the compatibility among nodes depends on the system size according to the compatibility threshold, $\mathbb{C}_i$, given by,

\begin{equation}
	\mathbb{C}_i=\frac{d}{\tau},
	\label{eq1}
\end{equation}

\noindent where $d$ is a constant called \textit{compatibility distance} between node characters, which is determined  by a certain probability density function; and $\tau$ represents the size $N$ of the system at that moment. The dependence of $\tau$ makes the probability of linking remains constant over time. This means that, as more nodes appear in the network, the new ones arrived to the system should be more like to the oldest nodes to which the link occurs. This perspective has major implications, which will be seen in the last section of this work.

\begin{figure}[htp]
\begin{center}
  \includegraphics[width=3in]{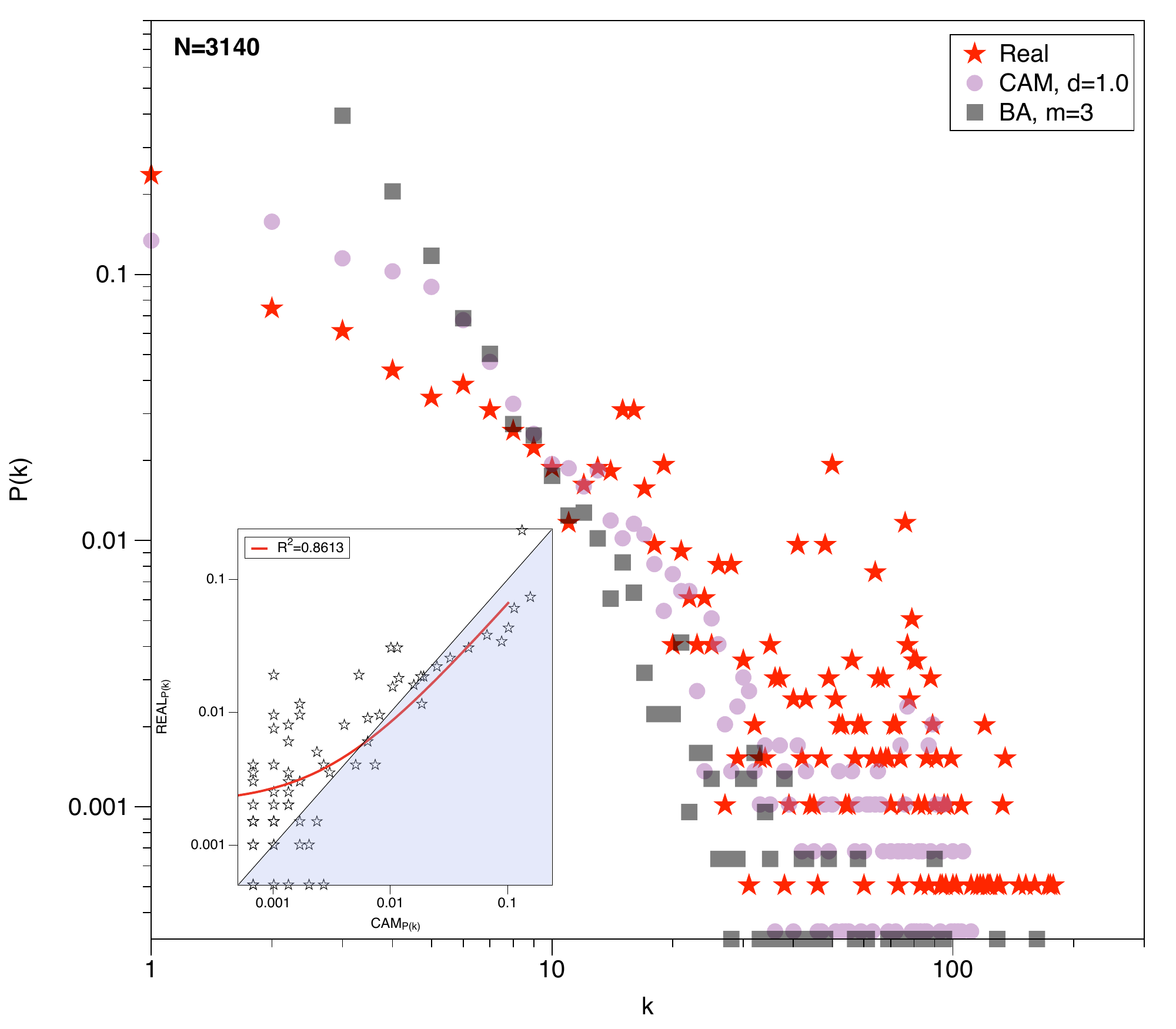}\\
  \caption{Node degree distribution, $P(k)$, for the real network (red stars) and two stochastic models: CAM (purple dots) and BA model (grey squares). Bottom inset shows the correlation between $P(k)$ of real network and those generated by the CAM.}
  \label{fig5}
  \end{center}
\end{figure}

Figure \ref{fig5} shows the degree distribution, $P(k)$, observed both in the real power network and in the two models previously described, BA model and CAM. As can be seen, both models generate a network with heterogeneous  connectivity, with many  poorly-connected  nodes and few \textit{hubs}. However, the CAM more accurately reproduces the real connectivity distribution. The  BA model  gives poorly  connected  people  a higher  probability in comparison with the real network;  moreover, the scaling exponent is stronger, which  denotes less connectivity heterogeneity  with respect  to the real network.  The inset  of Figure \ref{fig5} shows the  high correlation of finding people with $k$ connections  in the real and CAM networks.

CAM not  only closely reproduces  the  connectivity distribution, it also reproduces the  high  transitivity of connections $\hat{C}$ and  the  mean connectivity $\hat{k}$ better than the  BA  model (Table \ref{tab2}). The  BA model does have a fairly faithful reproduction of  the  average  shortest path length $\hat{l}$, whereas CAM overestimates this metric.

\begin{table}[h]
 \caption{Models comparison. Number of nodes, links, scaling exponent $\gamma$, transitivity of connections  $\hat{C}$, average shortest path length $\hat{l}$, mean connectivity  $\hat{k}$, percentage of nodes that compose the giant component $GC$ and the degree assortativity  $r$, for the real network and two stochastic models: CAM and BA model.}
\begin{tiny}
\begin{tabularx}{250pt}{llllllll}
\hline
Network			 		& Links 		& $\gamma$  		& $\hat{C}$ 			&  $\hat{l}$		&  $\hat{k}$		&  GC (\%)	&  $r$\\
\hline \\ 
Real 					& 17353		&-1.5			& 0.622 				& 3.69			& 17.0			& 60.0	& 0.38 \\
CAM  		& 13741.4 $\pm$ 593		&-1.6			& 0.596 $\pm$ 0.005	& 7.65 $\pm$ 0.34	& 8.8 $\pm$ 0.4	& 60.6 $\pm$ 2.4	& 0.88 $\pm$ 0.01 \\
BA 			& 9411.0 $\pm$ 0		&-2.0			& 0.013 $\pm$ 0.001			& 3.88 $\pm$ 0.007			& 5.0 $\pm$ 0			& -		& $\sim$0.0 \cite{Asso}\\
\end{tabularx}
\label{tab2}
\end{tiny}
\end{table}

Another  interesting aspect of the CAM is that it is capable of generating  a graph with a giant component, as occurs in the real network  where the entire  graph is composed by isolated  subgraphs and one giant network that groups over 60\% of the people.  In the BA model the resulting network graph has a unique connected  component. Moreover,  the  CAM  can  reproduce  the  assortativity  \cite{Asso} ($k$-correlation) observed in  the  real  network.    The  last  was  expected  in  the  CAM  because  the compatibility mechanism  generates  a  homophilic  network, however, the observed homophily in the real network is not trivial. Highly connected  people in the  power network  are preferably  linked  together, while those  with  few connections  are  linked  with  others with few connections.

It is interesting to note what  happened with  the  attribute of social power $P$. Well, the assortativity per person attribute ($p_i$)  is also positive $r_p$=0.11,  which means  that \textit{powerful people} (according  to our definition)  tend also to connect to powerful people and people with low power are linked to less powerful people.

\begin{figure}[htp]
\begin{center}
  \includegraphics[width=3in]{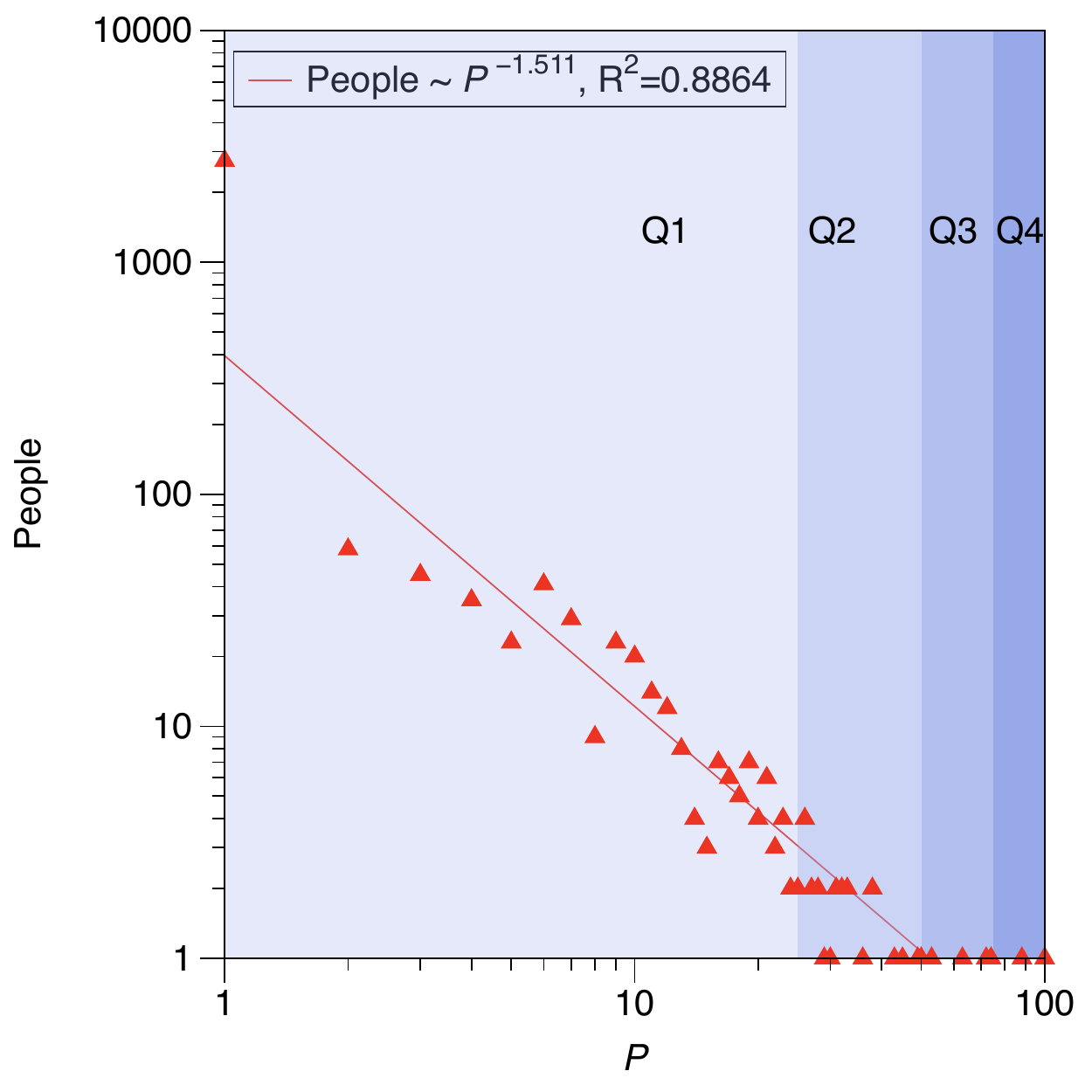}\\
  \caption{Number of people having social power $P$ in the network. The power was normalized and divided in quartiles (Q).}\label{poder}
  \end{center}
\end{figure}

Due to the  fact that, according to the Poderopedia power connection data, power $P$  seems to be strongly  correlated with  node degree, it is not surprising that the distribution of $P$ also follows a power law with an exponent similar to the $P(k)$ distribution (Fig. \ref{poder}). Using the same fit test as of section \ref{chile_net}, we can observe that the power $P$ also scales giving to 1\% of the population of the network (31 people) 31.7\% of the total $P$ power\footnote{Coincidently, the  percentage of income  that the top  1\% of Chileans accumulates is  30.5\% \cite{Lopez}.}. Approximately 2\% of the population has a half of power of the system. Notice that the most powerful quartile is composed of only 2 people, a clear social power concentration.

In order to explore the connections between \textit{powerful people}, and to validate the compatibility attachment model in this context, we use the distribution of power $P$ to establish a  ``nucleus of power'', defined as the top 20 politicians and businessmen to respective  positions  in the $P$ rank (40 people = 1.2\% of the population). Using the information of each entity stored in Poderopedia data, we selected only people that have an exclusively political or busisness ``career''. However, it is necessary to say that in spite of the result that shows an inhomogeneous distribution of $P$, the concept of ``nucleus of power'' it is not totally correct; this is because there are many nuclei at different scales, a \textit{Polyarchy} according to R. Dahl \cite{Dahl}. However, we will use this concept throughout the work to refer to the group of most powerful people.

\begin{figure}[htp]
\begin{center}
  \includegraphics[width=3in]{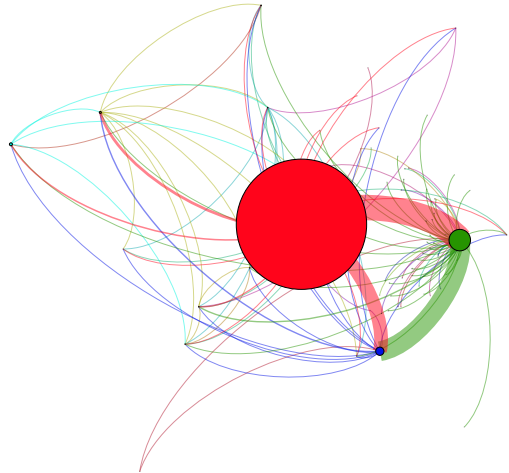}\\
  \caption{Network of Chilean Power. Layout as Fig. \ref{fig1} with people of the same community ``collapsed'' into a macro-node labeled according the most representative people group detected. Size of the node represents the amount of social power $P$ contained.}
  \label{fig6}
  \end{center}
\end{figure}

The same meaning of ``nucleus of power'' was used by W. Mills \cite{Mills} to refers to the most powerful people in the United States. Mills understood that it is not merit which determines the possibility of belonging and inclusion, but rather trust, a subjective factor that is related to the attitude of commitment to the defense of that position of power. It's what Michels understood as the ``iron law of oligarchy'' \cite{Michels}. The nucleus of power, in our view, is pierced  by  these kinds  of attitudes, which represent identity links, commitments supporting domain positioning, and optimal gratification involving rewards. The  remarkable thing  about  this is that this  phenomenon explained  by political  sociology can  be observed  in our analysis.

It is not surprising  that these forty people are located in the three  principal  communities. However, their  distribution on these  modules  is different.  For  example,  top  businessmen  belong to the  red and  blue communities  (50\% each one), while 65\% of the top politicians  belongs to the green community,  25\% to the red, and the rest to the blue. Figure \ref{fig6} show the same representation of the network of Figure \ref{fig2} but with communities collapsed into macro nodes labeled according to their principal type of top entities. It can be observed that the red community contains not only 50\% of the top businessmen and 25\% of the top politicians, but also the most powerful people (the size of the (macro) node represents the amount of power $P$ contained in it).

\section{An interpretation from sociological theory of Power and Elites}

Sometimes  it  seems more  interesting to  ask  \textit{why} a person  obeys  another, rather than trying  to explain  \textit{how} someone has power over another.  Political  sociology assumes this question with the understanding that, in this relationship, the search for ``prompt  and sincere'' obedience seems to be essential \cite{Weber}. Therein  lies legitimacy, a dimension  shared by all political  action.

Max Weber  and his followers (including  theorists who studied Elites, such as G. Mosca, W.  Pareto, W.  Mills and  R.  Michels)  have  built  a vast  analytical interpretation, which we delve into below (for details  of source theory,  see Table  \ref{matrix_nucleo}). Classification  of power and  the phenomenological  categories  explaining  it  are a powerful analytical tool box that we have  applied  in an initial  attempt to understand exactly what  is involved in the  interpretation of the  connections  between powerful people in Chile. These include the \textit{factors of domination} adduced  by G. Mosca (\textit{organizational skills} and \textit{capacity to create political formula}); the laws from R. Michels (\textit{specialization law}, \textit{law of psychological metamorphosis of the leader} and \textit{the iron law of oligarchy}), on which positivism  is based; the  \textit{psychological residues} left by W. Pareto, which defines permanence, combination, and  the  recurring  phenomenon of the \textit{circulation of elites}; and the analysis of the \textit{nucleus of power}, by W. Mills.

\begin{table}[htp]
 \caption{Conceptual categories of Power and Elites used in this work. Principal referent for each category and typologies associated. \label{matrix_nucleo}
}
\begin{tiny}
\begin{tabularx}{250pt}{XXX}
\hline 
Categories & Author &Typologies\\
\hline  
Power & M. Weaver \cite{Weber,Bendix,Aron,Ritzer} &Economic,  Coercive, Ideologic\\
  
Legitimate Domination& M. Weaver \cite{Weber,Bendix,Aron,Ritzer} & Charismatic, Traditional, Racional Legal\\
  
Nucleus of Power & W. Mills \cite{Mills} & Economic, Military, Politic \\
  
Objects of Situation & T. Parsons \cite{Sanchez,Ritzer} & Expected Roles, Internalized values, Recognition of affection \\
  
Domination Factors & G. Mosca \cite{Mosca,Mosca2} & Capacity to organize, Capacity to create political formula \\
  
Law of Oligarchy & R. Michels \cite{Michels} & Law of specialization, Law of psychological metamorphosis of the leader, Iron Law of Oligarchy\\
  
Psychological residues of Elites & W. Pareto \cite{Alonso1,Marmuc,WPareto} & Persistence residues, Residues of combination \\
  
Interests of Domination & R. Dahrendorf \cite{Ritzerb}& Latent interests, Manifests interest \\
  
Pluralistic Society &  W. Kornhauser \cite{Korn}& Accessibility of elites, Availability of non-elites\\
\end{tabularx}

\end{tiny}
\end{table}

\begin{figure}[h]
\begin{center}
  \includegraphics[width=3in]{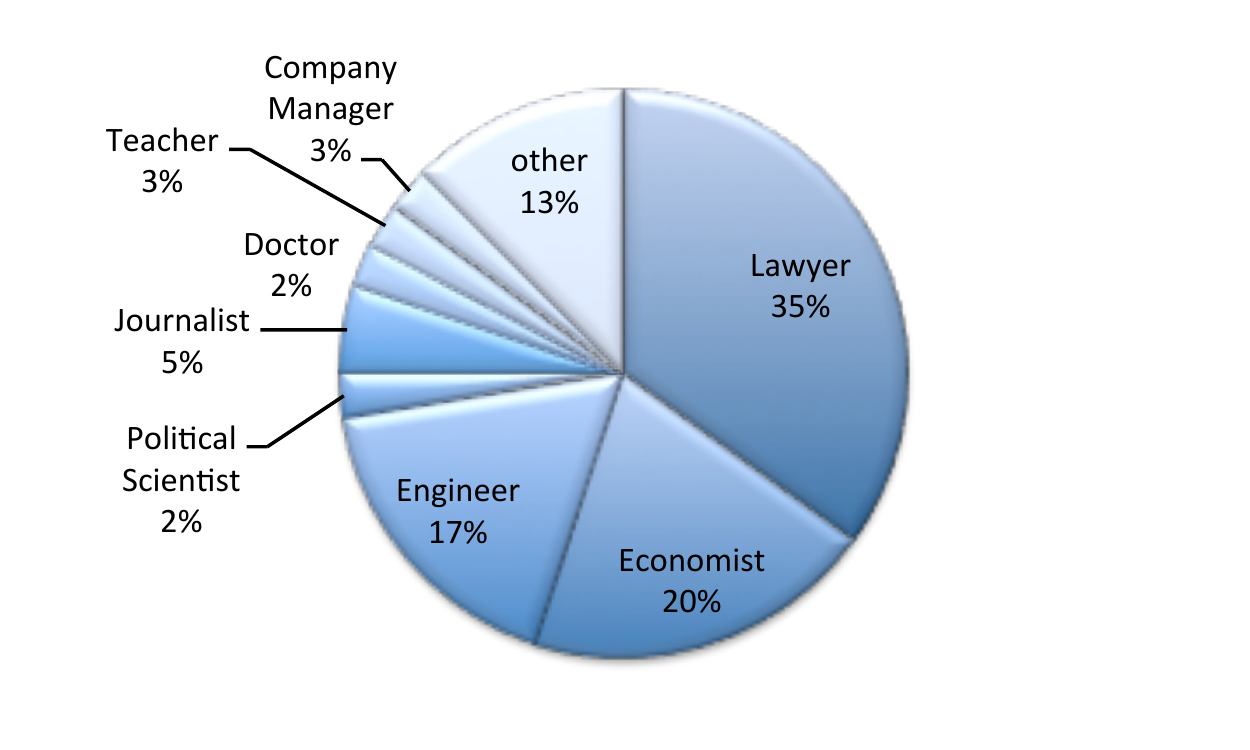}\\
  \caption{Distribution of professional training in the nucleus of power.}\label{fig7}
  \end{center}
\end{figure}

According  to  the  CAM,  the  power  network  of Chile  can  be interpreted as the  result  of a dynamic  population growth  process where relationships  between  people depend  on the  compatibility between  their  personal  characteristics.  In fact,  looking at  professional training and age characteristics within the ``nucleus of power'', we observed  that a  group with homogeneous attributes exists.  Figure  \ref{fig7} shows the distribution of professional training in the  nucleus,  where  70\% of the  people  are  lawyers, economists or engineers, with an average age of  60.7 $\pm$ 9.6 years. However,  as we might intuit, not  all 65 year old Chilean lawyers are powerful.  Deeper reasons  must  exist to explain this phenomenon.

Despite professional  training and  age  supporting the proposed compatibility thesis,  it was still possible that implicit, underlying  causes from Poderopedia's  infrastructure could have a measurable effect on relationship analysis  (see Table \ref{tab1}). In order to search  for these  causes, we explored paired connections  of nucleus nodes.  In this way we were able to circumvent one of the  major  weakness of modern  sociological theory,  which does not convincingly marry  interactions at the micro level with those of the macro-level \cite{Gra}. 

As many  possibles  paths  exist  between  politicians  and  businessmen  nodes in the  nucleus,  we analyzed  the  shortest paths, which are more robust for detecting  the  if a common view of society is held by both entities.  The inverse of this reinforces this choice; indeed Harary \textit{et al.} \cite{Har} noted that there  may be a distance  beyond which it is not feasible for a person to communicate with another because of the costs and distortions associated with the act of transmission. It is important to mention that our  analysis  makes use of the  original  Poderopedia graph  (\textit{i.e.}, before projection)  since not  only was the  detail  of personal relationships need, but also the relations  between  people and organizations.

\begin{figure}[htp]
\begin{center}
  \includegraphics[width=1.5in]{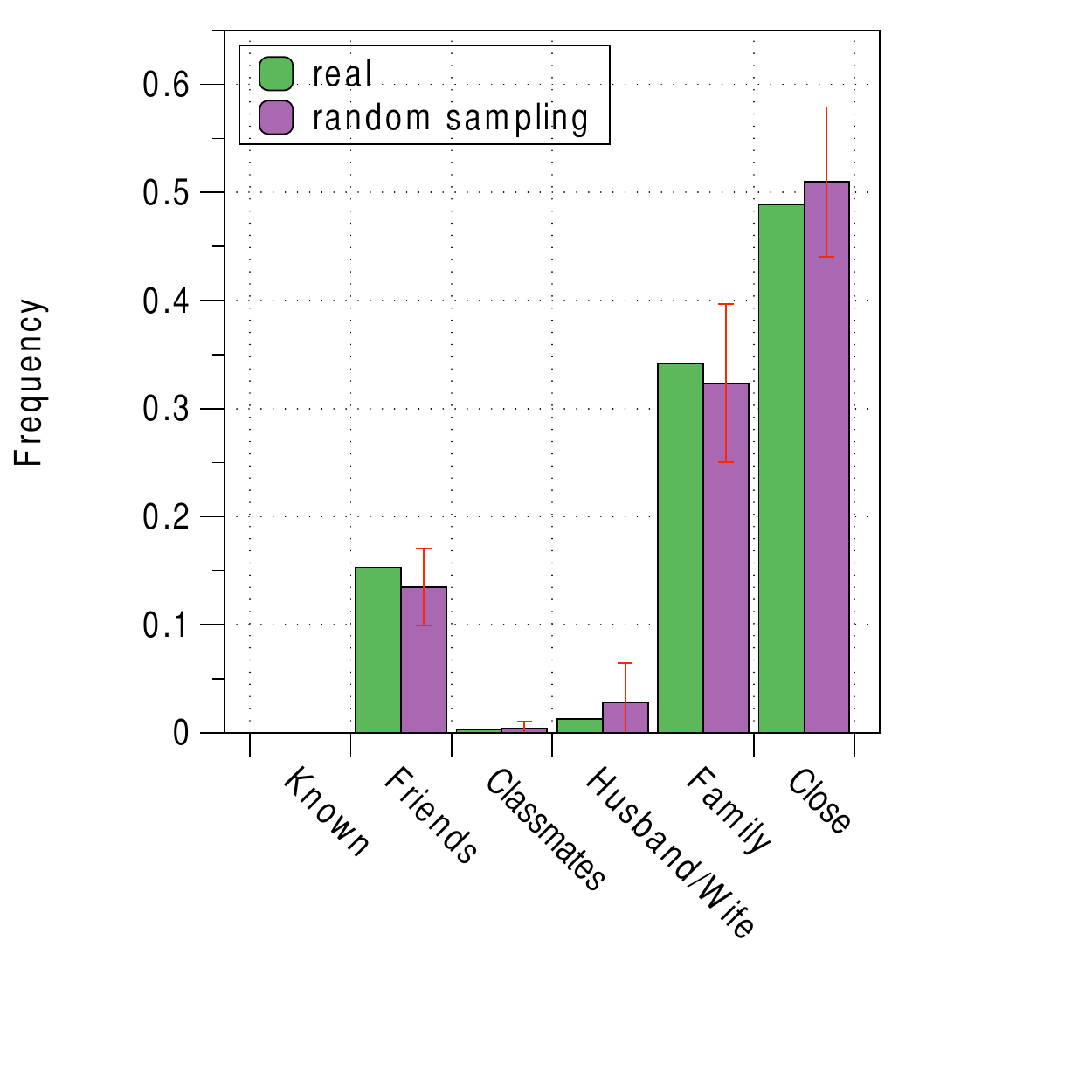}
  \includegraphics[width=1.5in]{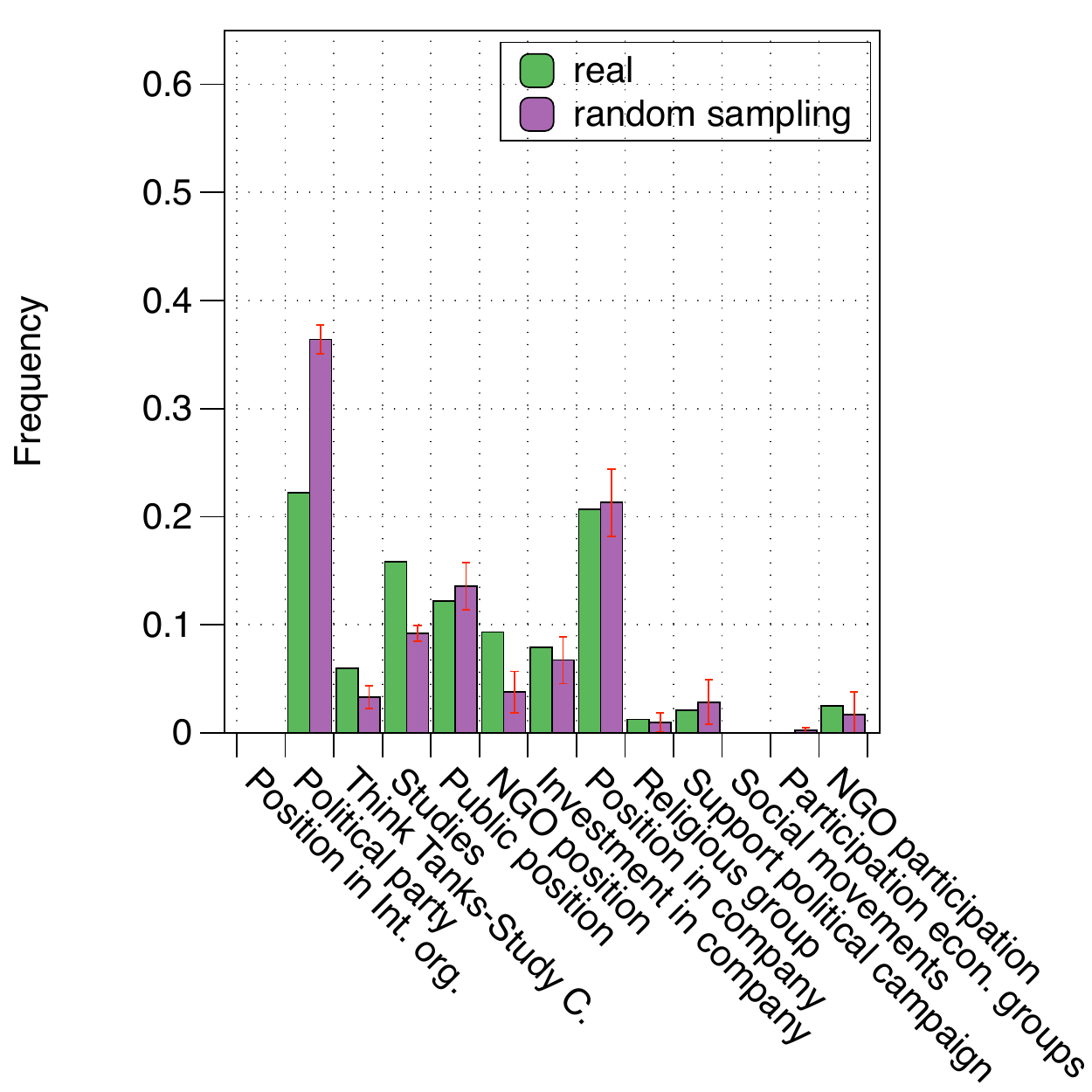}\\
  \caption{Frequency of links types detected in the paths computed for the nucleus of power. Person to Person (left) and Person to Organization/Companies (right). }\label{fig8}
  \end{center}
\end{figure}

The (shortest) path that connects two people was characterized by all the relation types and their path frequency.  Including  all the possible paths  computed, we obtained  a link type probability vector.  Since there were two link types considered  [person $\leftrightarrow$ person] and [person $\leftrightarrow$ organization], the vector was divided in both categories. Figure \ref{fig8} shows both vectors represented as bar charts where the height indicates the link type probability. 

As can be seen, 80\% of the relations, or social actions, in the nucleus of power are between close friends and family, followed by friends, in [person $\leftrightarrow$ person] relations. No significant difference can be observed  with the  paths  computed for random  sampling of those outside the top businessmen  and politicians.

On the other hand, considering [person $\leftrightarrow$ organization] actions, over 50\% of the relations inside the nucleus of power are between those belonging  to  the same political  party,  same place  of study,  or having  a position  in the same company.   For random  paths, the study  site loses importance in favor of political  party  relations.

What  we understand as relations or  social actions, is entirely related to the motivations that make it possible for someone to be linked to another, and any conditions that favor this link, including the intersubjective and homogeneous construction of ``world'' that  actors perform (shared codes). This is in agreement with \textit{rational motivators}, related  both to \textit{interests} and \textit{values}, as well as to \textit{traditions} and \textit{affections}.  This traditional classification given by M. Weber \cite{Runciman} was gathered and expanded upon by T. Parsons\cite{Parsons,Ritzer}. In order to give content to what happens along the paths that connect people, we derived \textit{attitudes}, \textit{expectations} and \textit{behaviors} associated with this typology of social action from the  computation of paths between  powerful people.

Table \ref{matrix1_used} shows the  interpretation that we used  for Poderopedia link  types as  according to theories of Weber and Parsons  \cite{Parsons,Ritzer}. The relationships contained in the Poderopedia data set represents the \textit{Situation of action},  according to Parsons, and reflects the scenario faced by an actor. This scenario can be evaluated according to the orientation modes that actors  take, which can be cognitive,  valorative,  or affective, in what  Parsons  calls ``motivational orientation''  \cite{Sanchez,Ritzer}. For cognitive orientation, we used the \textit{Expected Roles};  in the  valorative  orientation, the   \textit{Internalized Values};  and  in the  affective orientation, the  \textit{Recognition  of Affections} as  \textit{Objects of Situation}. Thus, our interpretation sees the actors as evaluating  their actions  according  to a rational  analysis  (cost / benefit);  an valorative analysis,  in order to estimate  the consequences  and effects of their  action  from the perspective of their  moral responsibility; and finally, an affective analysis,  in order to estimate the emotional satisfaction that the object will produce (desired).  Through case studies of the subjects, we emphasize the manifest  or latent character of each of these \textit{Objects of Situation}.

\begin{table}[h]
 \caption{Motivational Orientation associated to Poderopedia link types. Manifest (M) and Latent (L). [Source: own elaboration]. \label{matrix1_used}}
 \begin{tiny}
 \begin{tabularx}{250pt}{XXXX}
\hline 
&&Motivational orientation \tabularnewline
 & Cognitive orientation & Value orientation & Affective orientation  \tabularnewline
\hline 
&&Need Dispositions\tabularnewline

Situation & Expected Roles & Internalized Values & Recognition of Affections  \tabularnewline
\hline
\bf{[person $\leftrightarrow$ person]} \\ \hline
Known&$\bullet$(L)&&$\bullet$(M)\\ 
Friends&&$\bullet$(L)&$\bullet$(M)\\ 
Classmate&&$\bullet$(L)&$\bullet$(M)\\ 
Husband/Wife&&$\bullet$(L)&$\bullet$(M)\\ 
Family&&$\bullet$(L)&$\bullet$(M)\\ 
Close&&$\bullet$(L)&$\bullet$(M)\\ \hline

\bf{[person $\leftrightarrow$ organization]} \\ \hline
Position in International organisms&$\bullet$(M)&$\bullet$(L)&\\ 
Political party&$\bullet$(L)&$\bullet$(M)&\\ 
Think Tanks and Study Centers&$\bullet$(L)&$\bullet$(M)&\\ 
Studies&&$\bullet$(M)&$\bullet$(L)\\ 
Public position&$\bullet$(M)&$\bullet$(L)&\\ 
NGO position&$\bullet$(M)&$\bullet$(L)&\\ 
Investment in company&$\bullet$(M)&$\bullet$(L)&\\ 
Position in company&$\bullet$(M)&$\bullet$(L)&\\ 
Religious group&$\bullet$(L)&$\bullet$(M)&\\ 
Support groups to political campaigns&$\bullet$(L)&$\bullet$(M)&\\ 
Social movements&$\bullet$(M)&$\bullet$(L)&\\ 
Participation in economic groups&$\bullet$(M)&$\bullet$(L)&\\ 
NGO participation&$\bullet$(M)&$\bullet$(L)&\\\hline
\end{tabularx}

\end{tiny}

\end{table}

In addition, and given the interpretation framework,  paths connecting  two people contain  information about \textit{symbols} and \textit{meanings}. People act on the objects of their  world and interact with others  utilizing the subjective meanings that objects and people have, that is, from ``symbols''. In turn, ``meanings'' are the product of social interaction, mainly communication. Communication then  becomes essential  in the  constitution of both the  person  and  in the  social production of meaning.

\begin{figure}[htp]
\begin{center}
  \includegraphics[width=3 in]{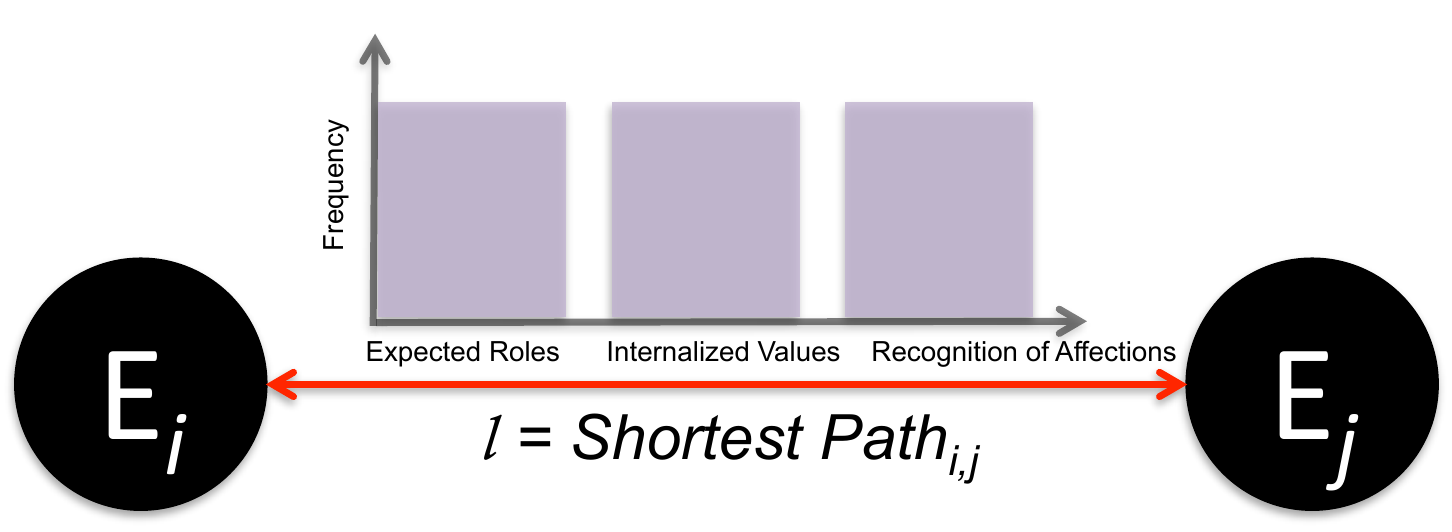}
  \caption{Path between entities $i$ and $j$. Length and amplitude defined.}\label{fig9}
  \end{center}
\end{figure}

A relevant point from sociology highlights that interaction (social action) is only possible when actors uniformly understand the senses and meanings, codes and symbols, of either party involved in the relation.  This point is also key to understanding the phenomenon that supports the construction of a ``nucleus of power''. As discussed throughout, the ``powerful'' tend  to be linked with themselves,  which does not solely happen  by a factor  of domination but  rather by one of likeness \cite{Mills}.

Figure \ref{fig9} shows an example of a proposed method including the interpretation rules discussed above. In this example, entity $i$ and $j$ are separated by $l$ entities denoting the length of the path. Here, we identify another   property of the  path:   its  amplitude.   Path  amplitude is defined as  the frequency  of the (for example) the three  motivational orientations:  Expected Roles, Internalized Values,  and  Affect  Recognition,   along  the  shortest path  between  two entities according  to Table \ref{matrix1_used}.

\begin{figure}[htp]
\begin{center}
  \includegraphics[width=1.5 in]{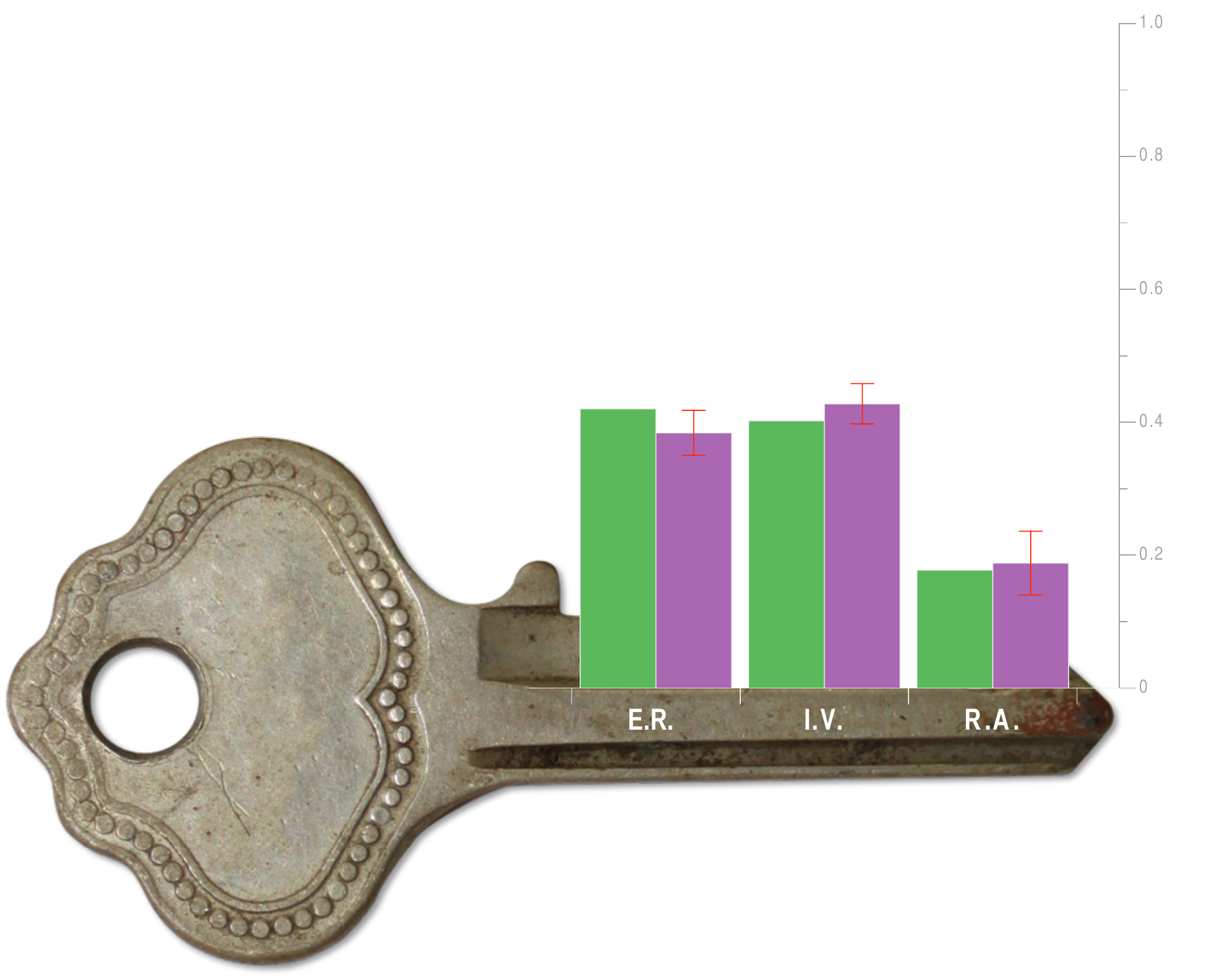}
   \includegraphics[width=1.5 in]{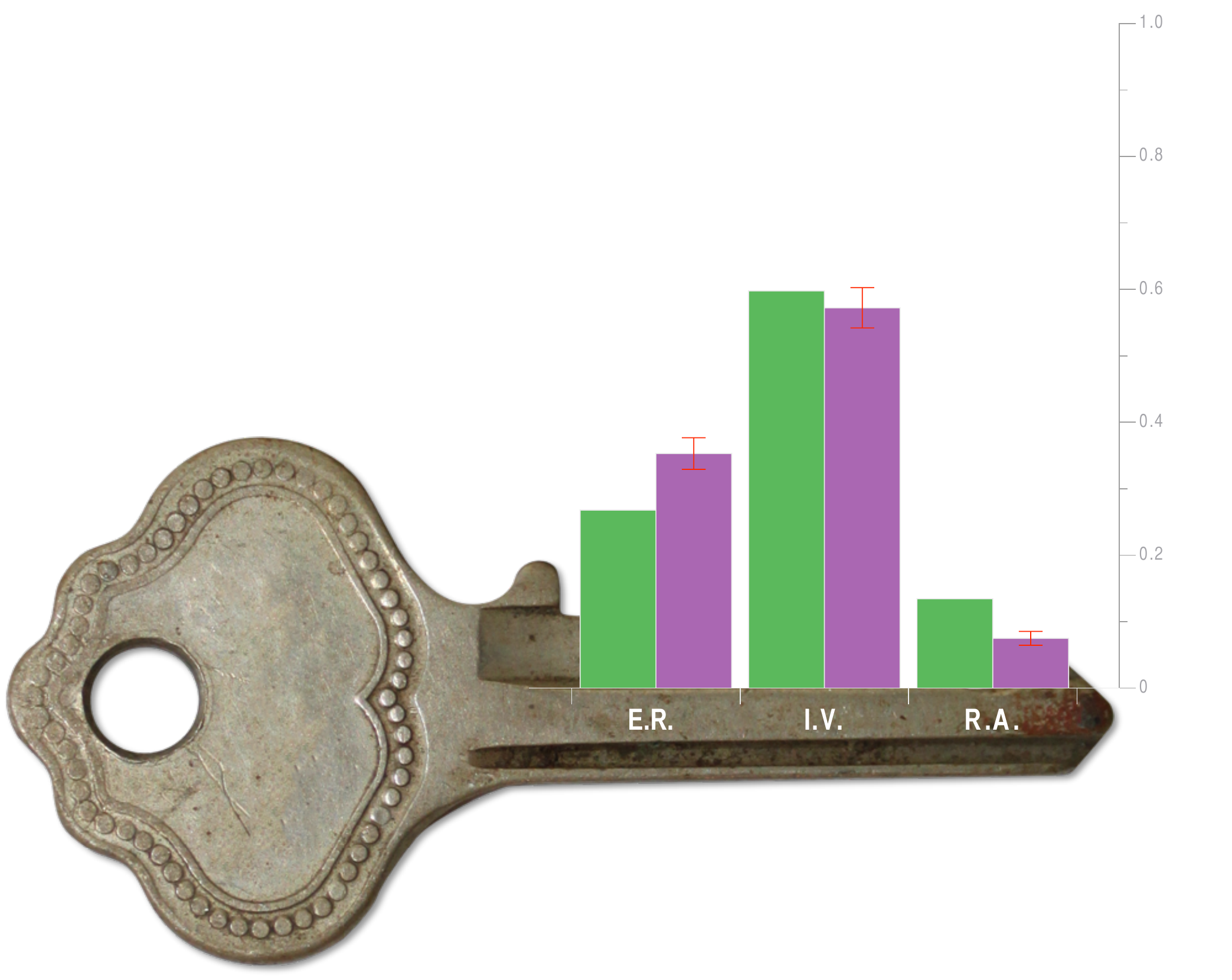}
  \caption{Keys of compatibility: Manifest Motivational Orientation (left) and Latent Motivational Orientation (right). According to Table \ref{matrix1_used}. Nucleus of power (green) and 3 random measures (purple). E.R.= Expected Roles, I.V.=Internalized Values, R.A.=Recognition of Affections.}
  \label{fig10}
  \end{center}
\end{figure}

This  amplitude can  be interpreted as a ``key'' or ``code'' that allows two  entities to belong to the nucleus, or simply be linked to another. In simple words, the ``key of compatibility''  is the deeper compatibility between  people that we found.

Figure \ref{fig10} shows the ``key of compatibility'' for nucleus members. The result suggests that the links between powerful people obey their Expected  Roles  and Internalized Values to a similar  magnitude. Affect Recognition seems to be less important. No difference was observed between nucleus relation  and random relations,  suggesting  a robust  result.   However,  the  analysis  of latent motivations shows that Internalized Values have a ``starring'' role followed by Expected Roles. The recognition  of affect also seems to be the less important.

\begin{figure}[htp]
\begin{center}
  \includegraphics[width=1.5 in]{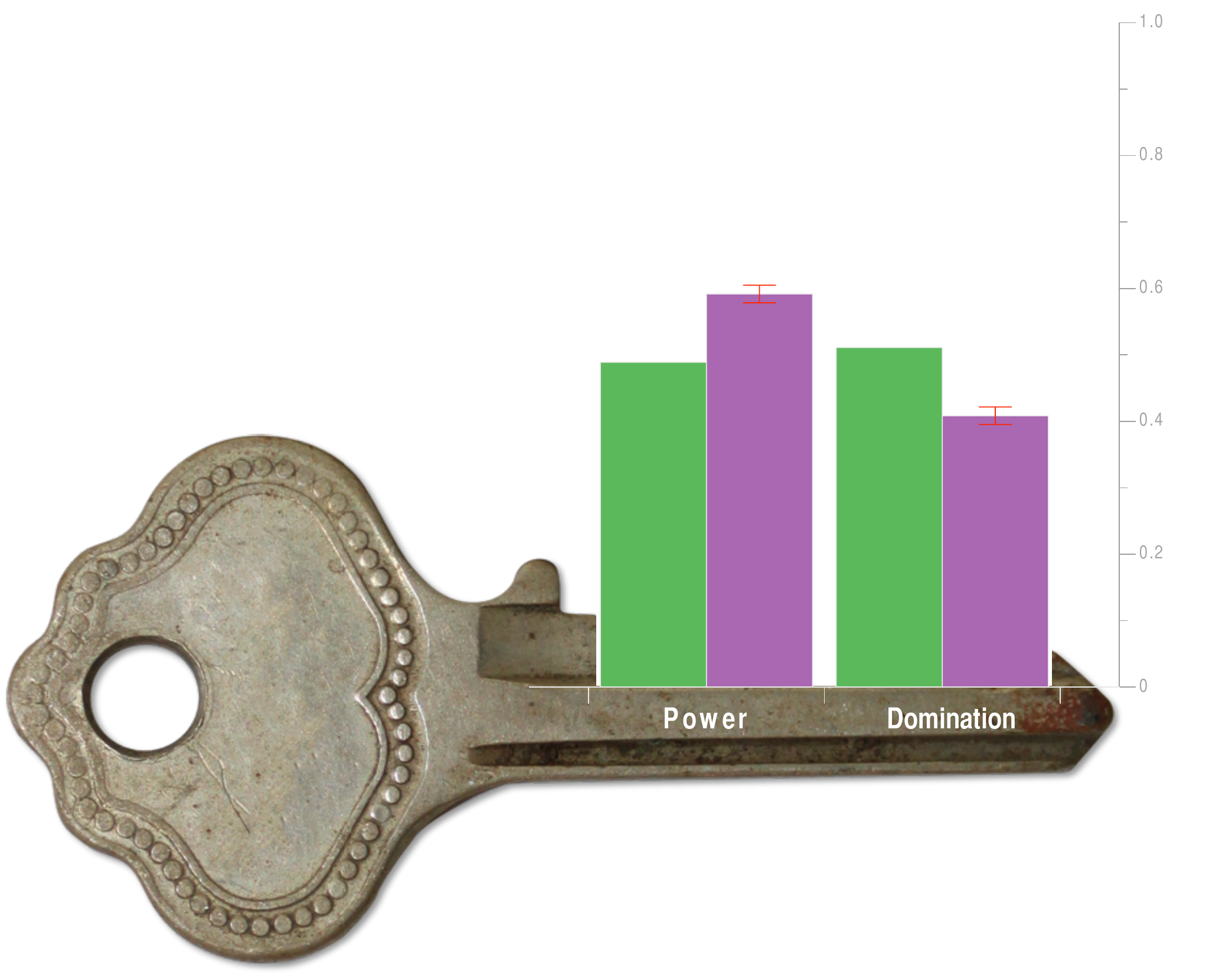}
   \includegraphics[width=1.5 in]{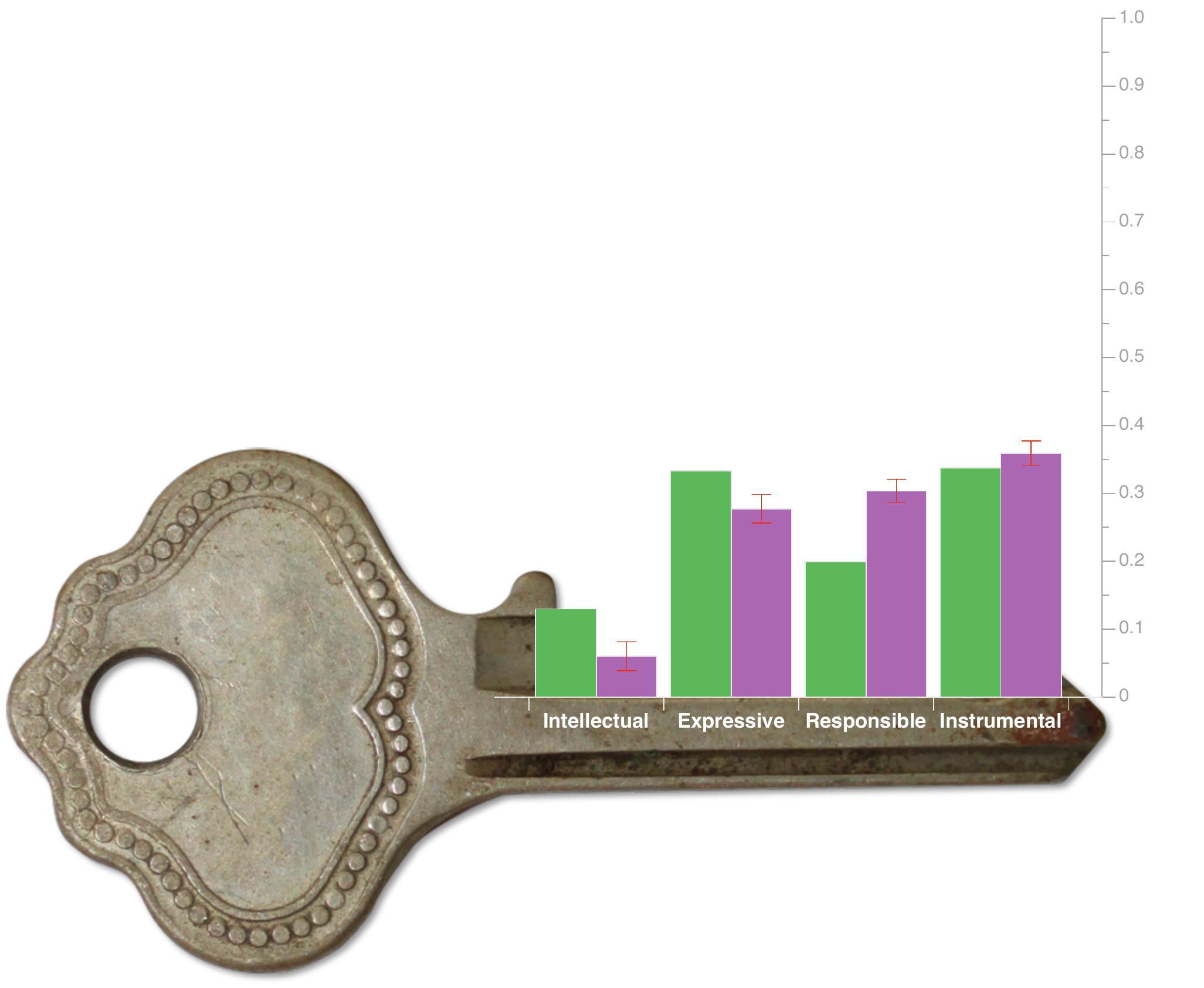}
  \caption{Keys of compatibility: Optimum of Gratification (left) and Resulting Basic Action Types (right). According to Table \ref{matrix1_used_B}. Nucleus of power (green) and 3 random measures (purple).}
  \label{fig14}
  \end{center}
\end{figure}

The previous results  can be further reinforced by taking the theory of Optimal Gratification into consideration, that is, by adding the analysis of vested interests of actors  and behaviors  associated  with the social action (Table \ref{matrix1_used_B}). In terms of motivation, the ultimate interest of any actor is to obtain the ``Optimum of Gratification'' as the best that can be obtained from the conditions given the set of existential needs and the set of possibilities. Thus, in order to designate the interests embodied in the \textit{Objects of Situation}, Parsons \cite{Sanchez,Ritzer} used the same as those of \textit{vested interests}. These interests are related to ``Power'' and ``Domination''. Additionally,  there  are resulting  behaviors  linked  to each one of the \textit{Objects of Situation} of Table \ref{matrix1_used} (cognitive/Intellectual-Instrumental; Valorative/Responsible; Affective/Expressive). Figure \ref{fig14} shows others ``keys of compatibility''  for the nucleus of power.

\begin{table}[h]
 \caption{Optimum of Gratification and Resulting Basic Action Types associated to Poderopedia link types. [Source: own elaboration].}
 \begin{tiny}
 \begin{tabularx}{250pt}{XXX|XXXX}
\hline 
 & \multicolumn{2}{c|}{Optimus of gratification} & \multicolumn{4}{c}{Resulting Basic Action Types}\tabularnewline
\hline 
 & \multicolumn{2}{c|}{Vested Interests} & \multicolumn{4}{c}{Behavior} \tabularnewline
\hline 
& \multicolumn{2}{c|}{Need Dispositions} \tabularnewline
Situation & Power & Domination & Intelectual & Expressive & Responsible & Instrumental \tabularnewline
\hline
Known, Friends, Classmate, Husband/Wife, Family, Close&&$\bullet$&&$\bullet$\\ \hline

Position in International organisms, Political Party, Think Tanks and Study Centers&&$\bullet$&$\bullet$\\ \hline

Studies&$\bullet$&&$\bullet$\\ \hline

Public position&$\bullet$&$\bullet$&&&&$\bullet$\\ \hline

NGO position or participation&&$\bullet$&$\bullet$&&&\\ \hline

Investment or position in company, Support groups to political campaigns&$\bullet$&&&&&$\bullet$\\ \hline

Religious groups&&$\bullet$&&$\bullet$&&\\ \hline

Social Movements&&$\bullet$&&$\bullet$&$\bullet$&\\ \hline
Participation in economic groups&$\bullet$&&$\bullet$&&&\\\hline

\end{tabularx}
\end{tiny}
\label{matrix1_used_B}
\end{table}

As can be seen in the left ``key of compatibility'' of Figure \ref{fig14}, actions of powerful people are related to expected ``Power'' and ``Domination'', both in a similar magnitude. Among the  randomly sampled (less-powerful people) gratification seems to be related  to power more than  domination. With respect  to the behaviors  derived  from the \textit{Situation of action}  (right ``key of compatibility''), the  results  suggest  that most  of the  resulting behaviors  are ``Expressive'' and ``Instrumental'', followed by ``Responsible'', and much further along, ``Intellectual''.  A similar pattern is seen in the random sampling  of less-powerful persons.

\subsection{Businessmen \textit{vs} Politicians}

\begin{figure}[htp]
\begin{center}
  \includegraphics[width=1.5in]{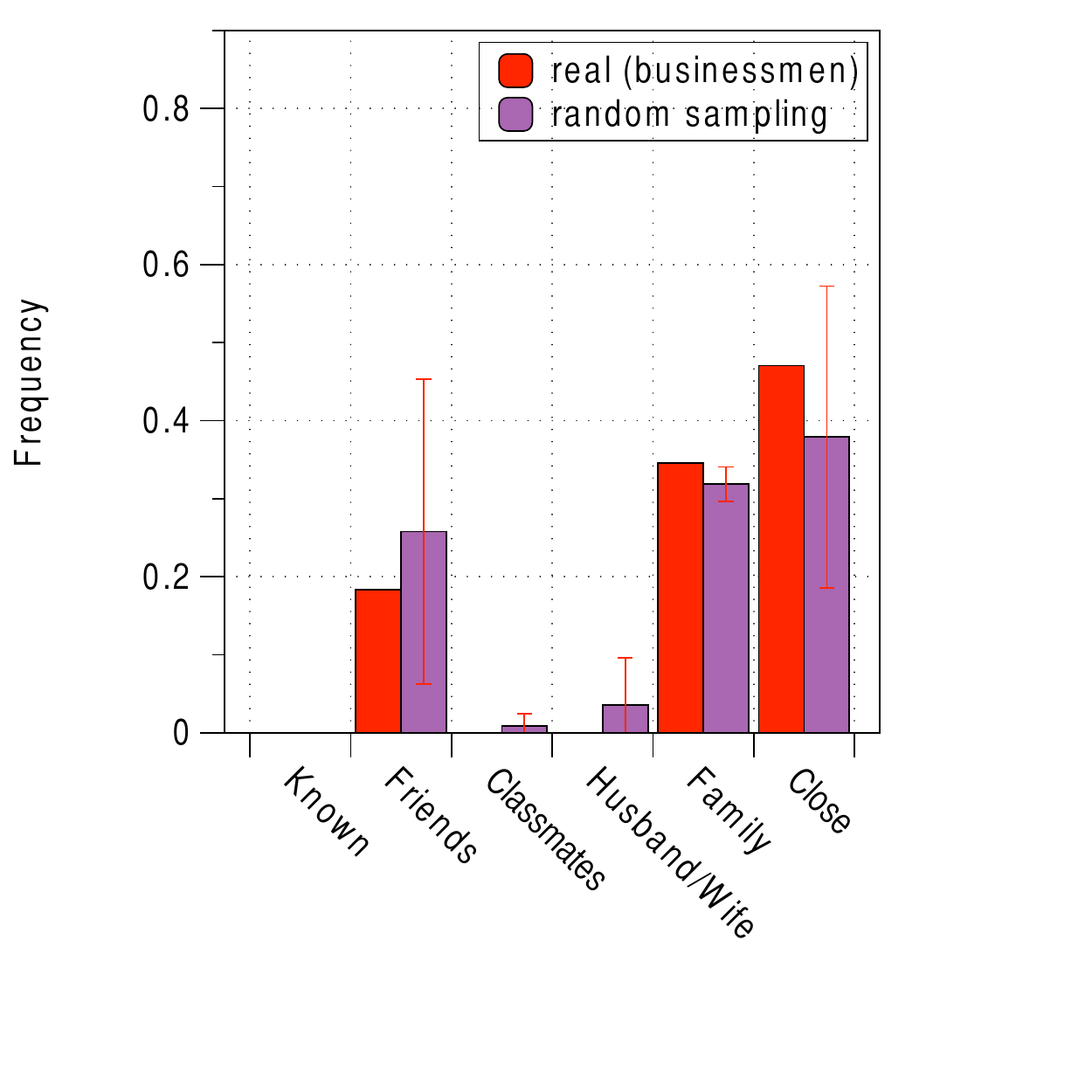}
  \includegraphics[width=1.5in]{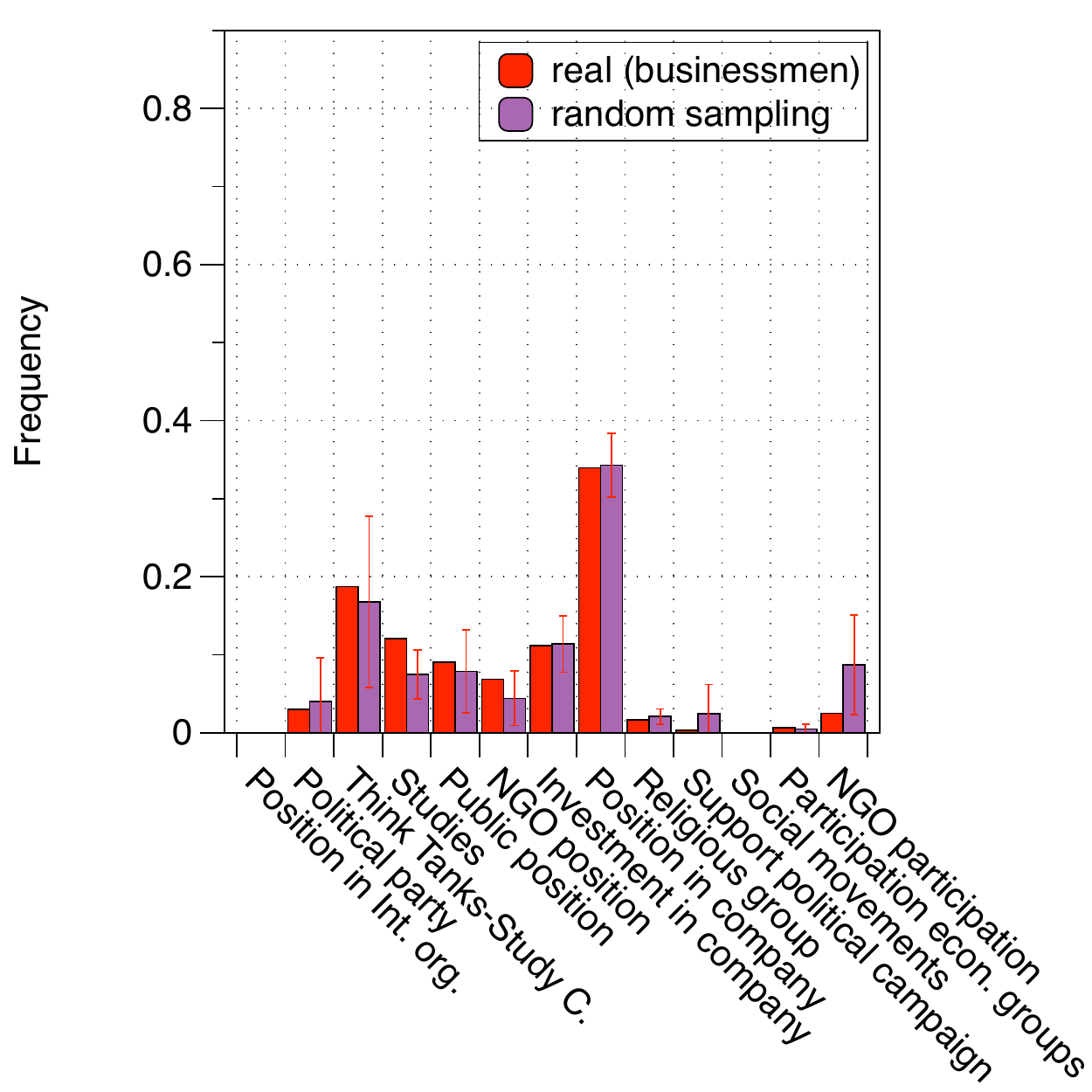}\\
   \includegraphics[width=1.5in]{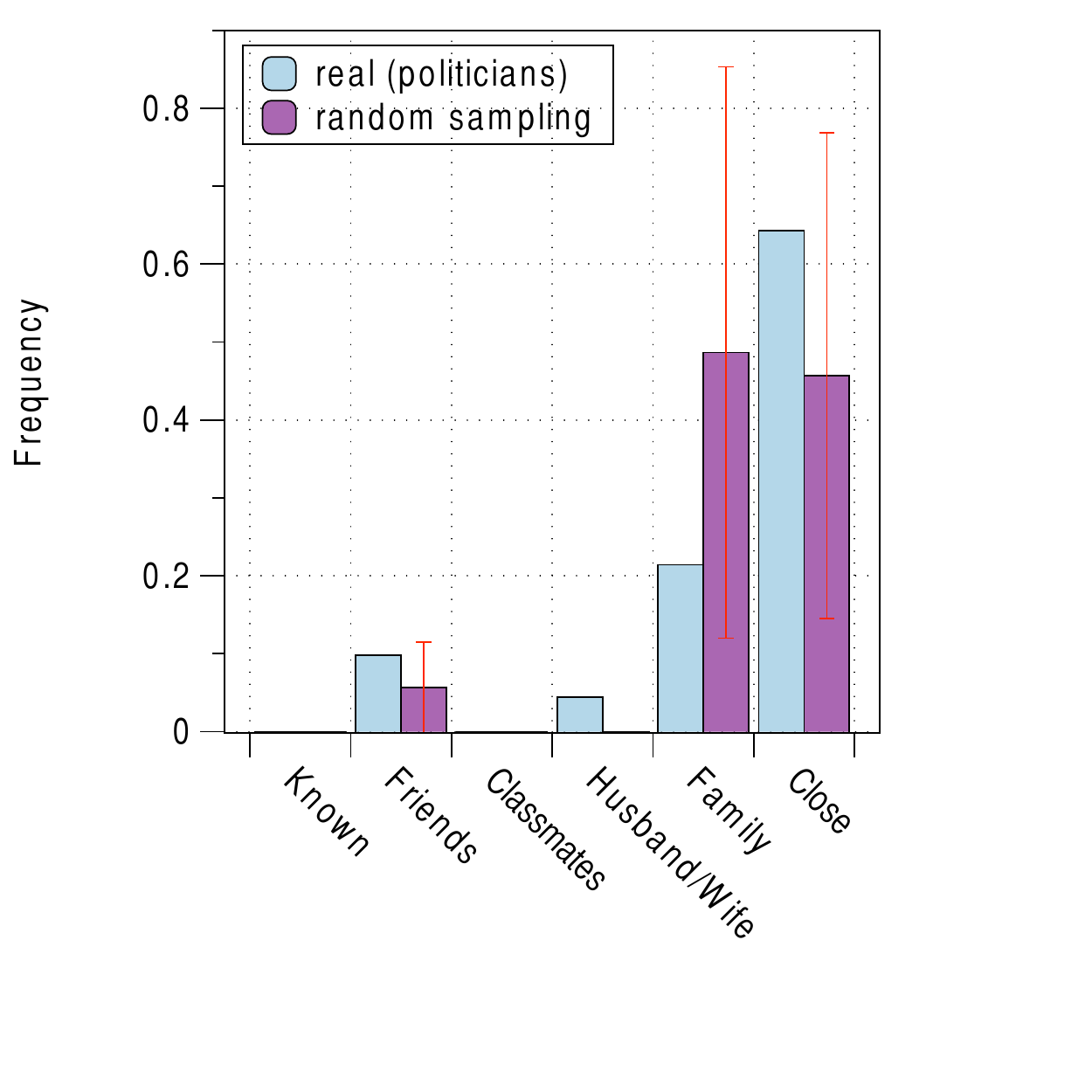}
  \includegraphics[width=1.5in]{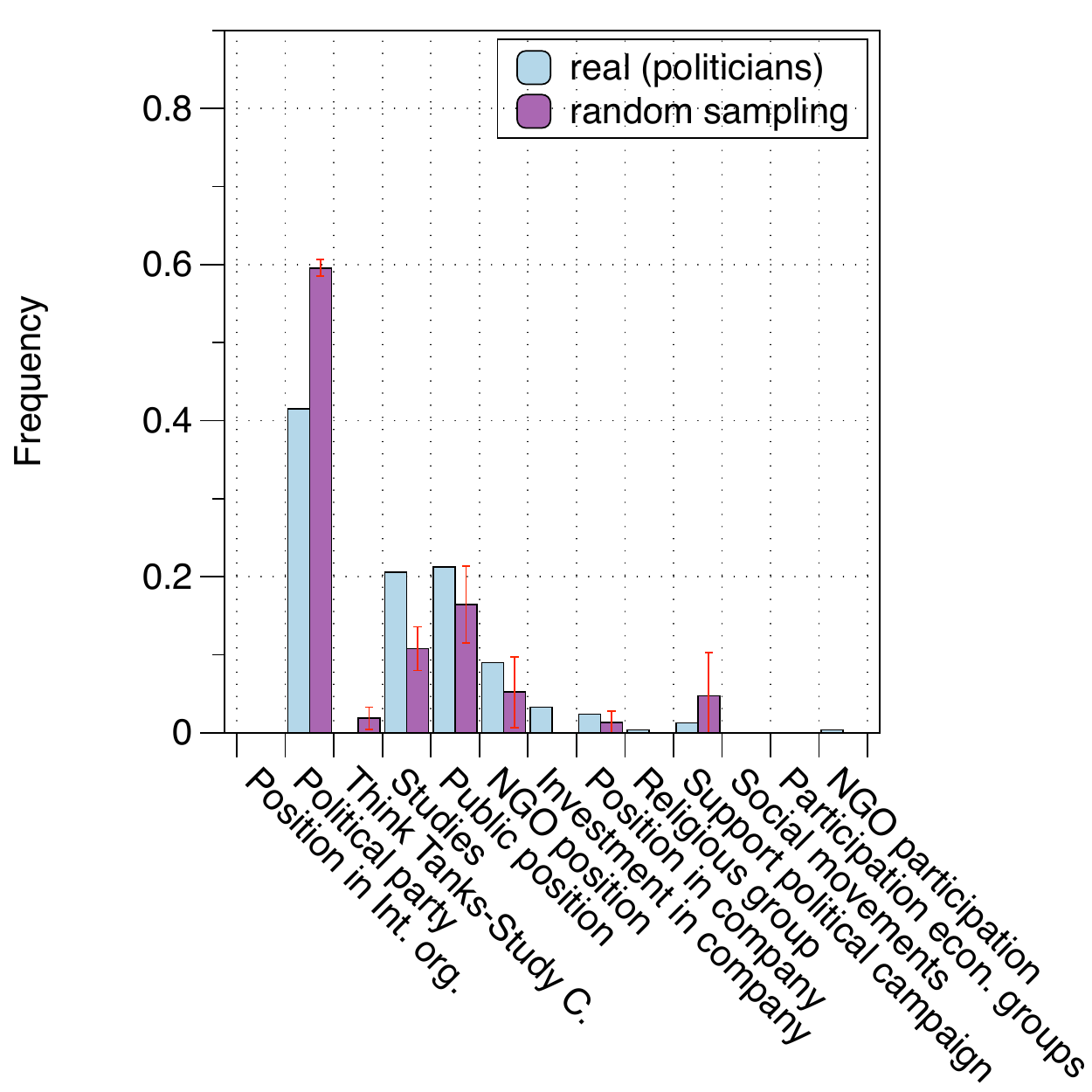}
  \caption{Businessmen relations: Person to Person (left top) and Person to Organization (right top). Politicians relations: Person to Person (left bottom) and Person to Organization (right Bottom). Frequency of links detected in the paths computed. Businessmen (red), politicians (cyan) and 3 random sampling (purple).}\label{fig11}
  \end{center}
\end{figure}

According to \cite{chilepoder}, power in Chile is concentrated in two areas: business and politics. In contrast, symbolic and social power represent less than 30\%. In order  to explore the  relations  inside the two main groups  and  to search for the  same  signals  of relation  detailed  in the  previous  section,  we analyzed  the shortest paths  between  businessmen  and  politicians. Figure \ref{fig11} shows the  vector of probability for link types that describe the  relationships  between top  people in both groups.

As can be observed, for [person $\leftrightarrow$ person] context, the social actions  between businessmen  are characterized by close and familiar  relations.  However, for the case of politicians,  most  of the  relations  are between  close people, while family or friend relations  are infrequent.  In the  random politician  paths, interestingly,   we observed  that family ties seem to play a much more important role, denoting  that these kind of links are less frequent  between  politicians  within the nucleus.  

The [person $\leftrightarrow$ organization] context yielded  somewhat expected results. The  most probable  relation  type  for businessmen  is their  participation in a company,  while for politicians,  belonging to a political  party  seems to be the  most  probable,  especially in random  paths. However,  this context  exposes something  else, namely, that paths  for businessmen  are more diverse than  politician paths.

\begin{figure}[h]
\begin{center}
  \includegraphics[width=1.5 in]{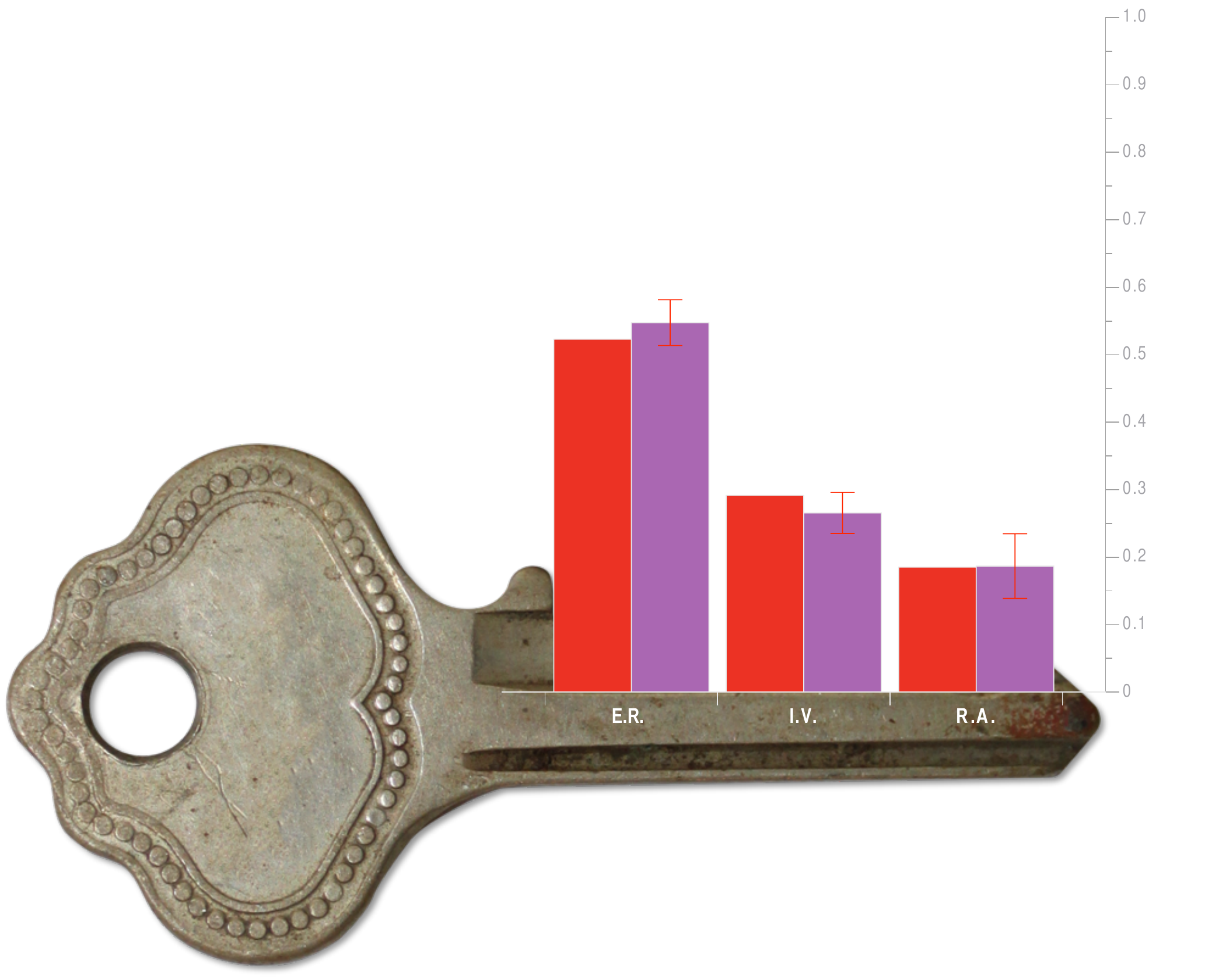}
   \includegraphics[width=1.5 in]{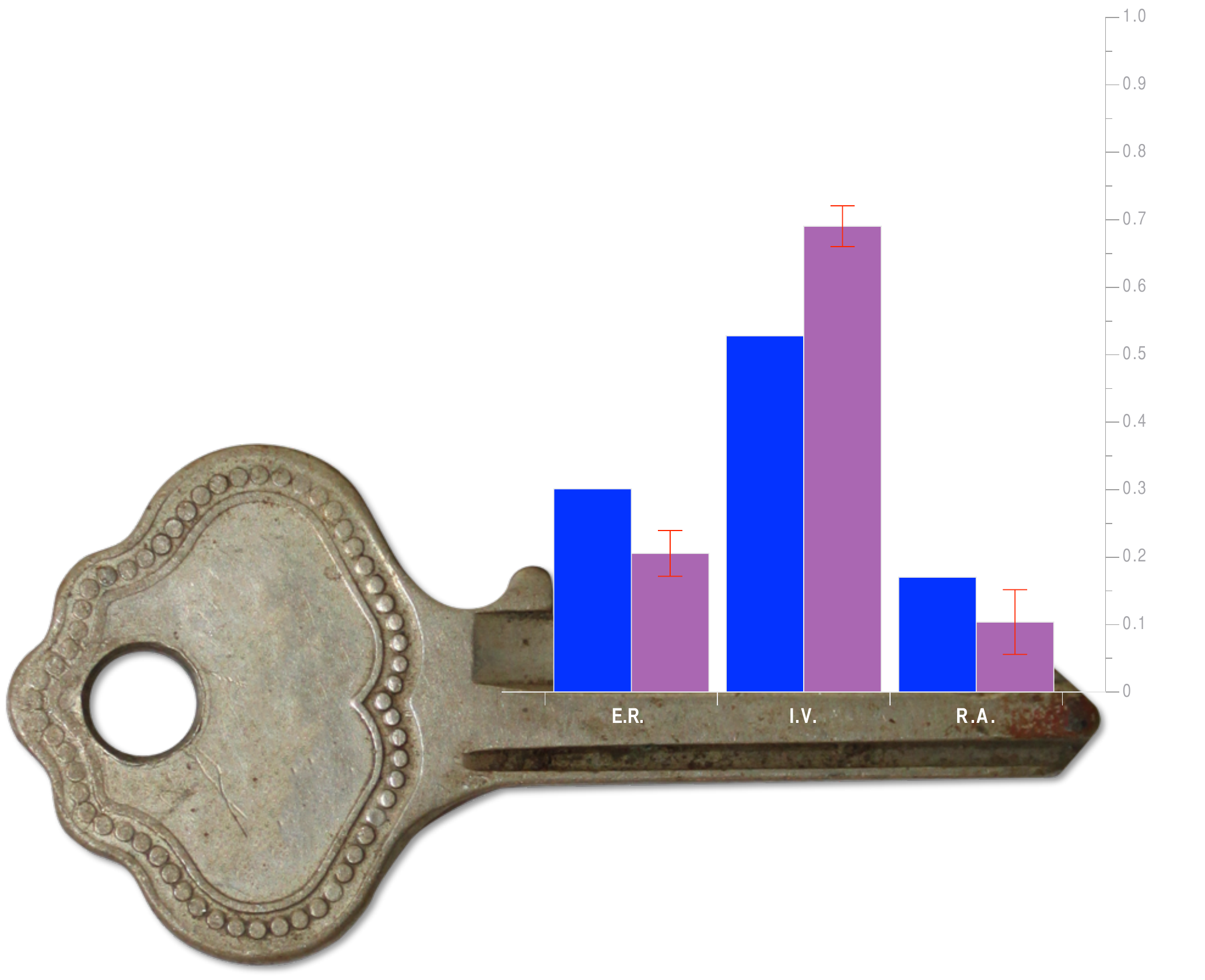}
   \includegraphics[width=1.5 in]{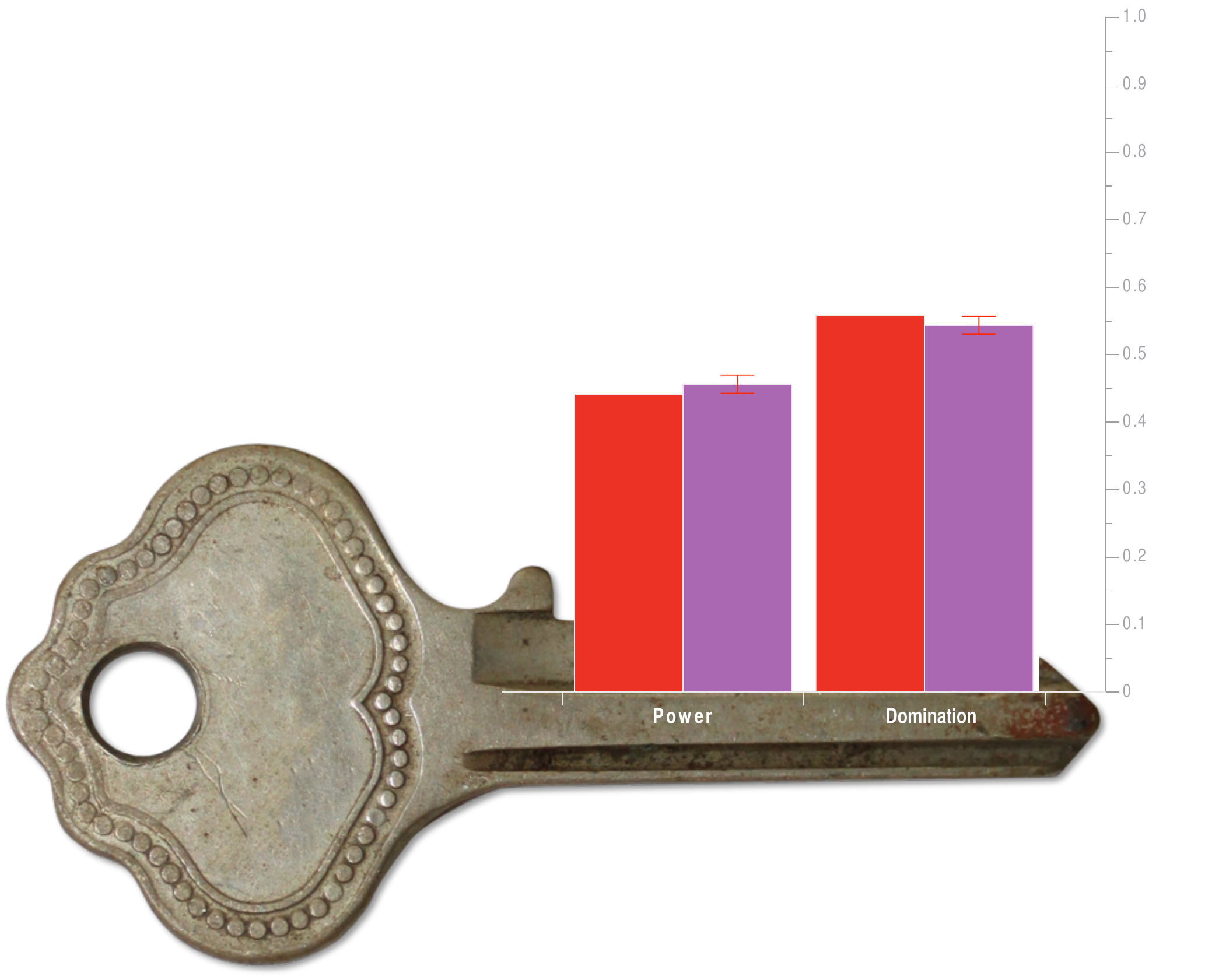}
   \includegraphics[width=1.5 in]{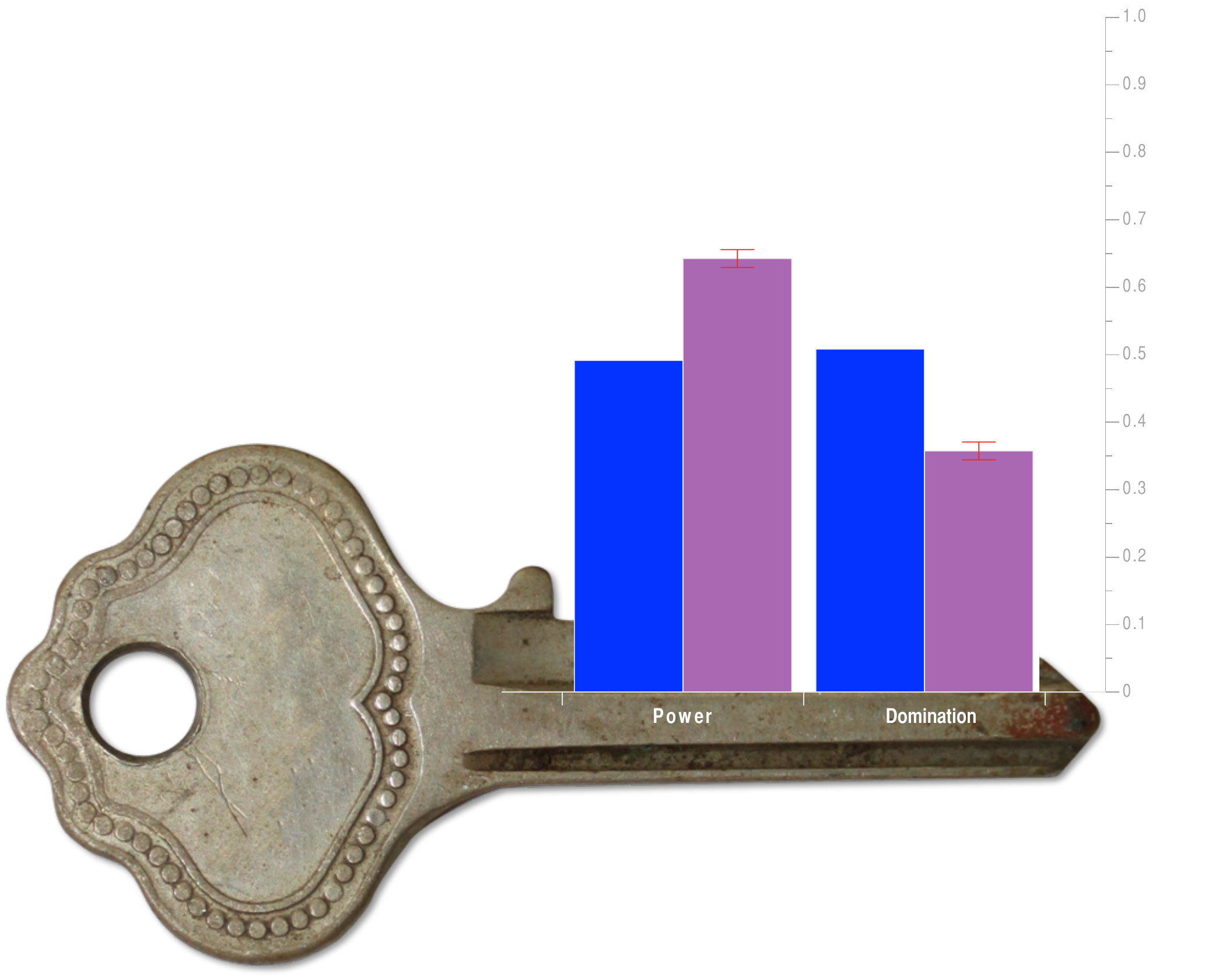}
\includegraphics[width=1.5 in]{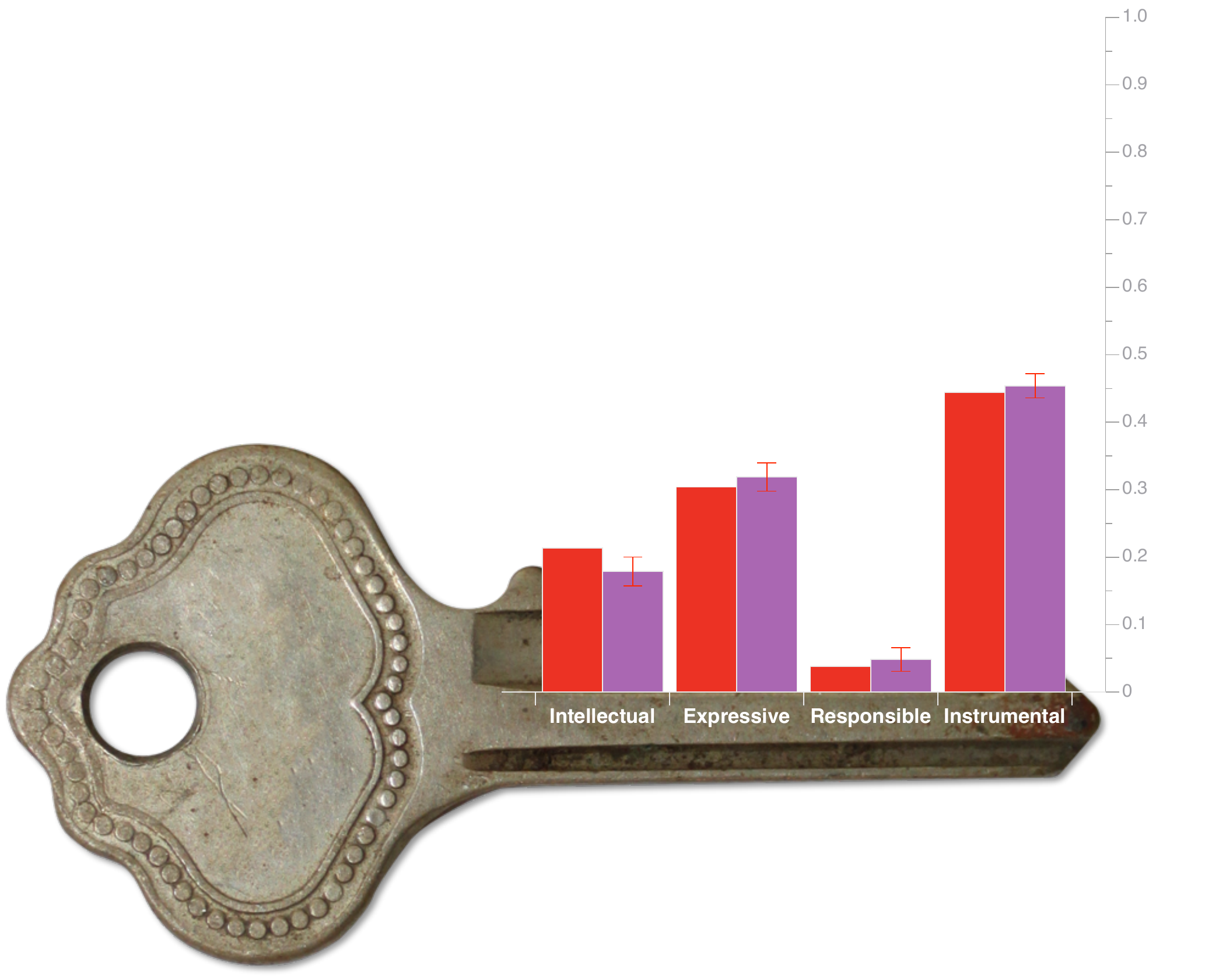}
   \includegraphics[width=1.3 in]{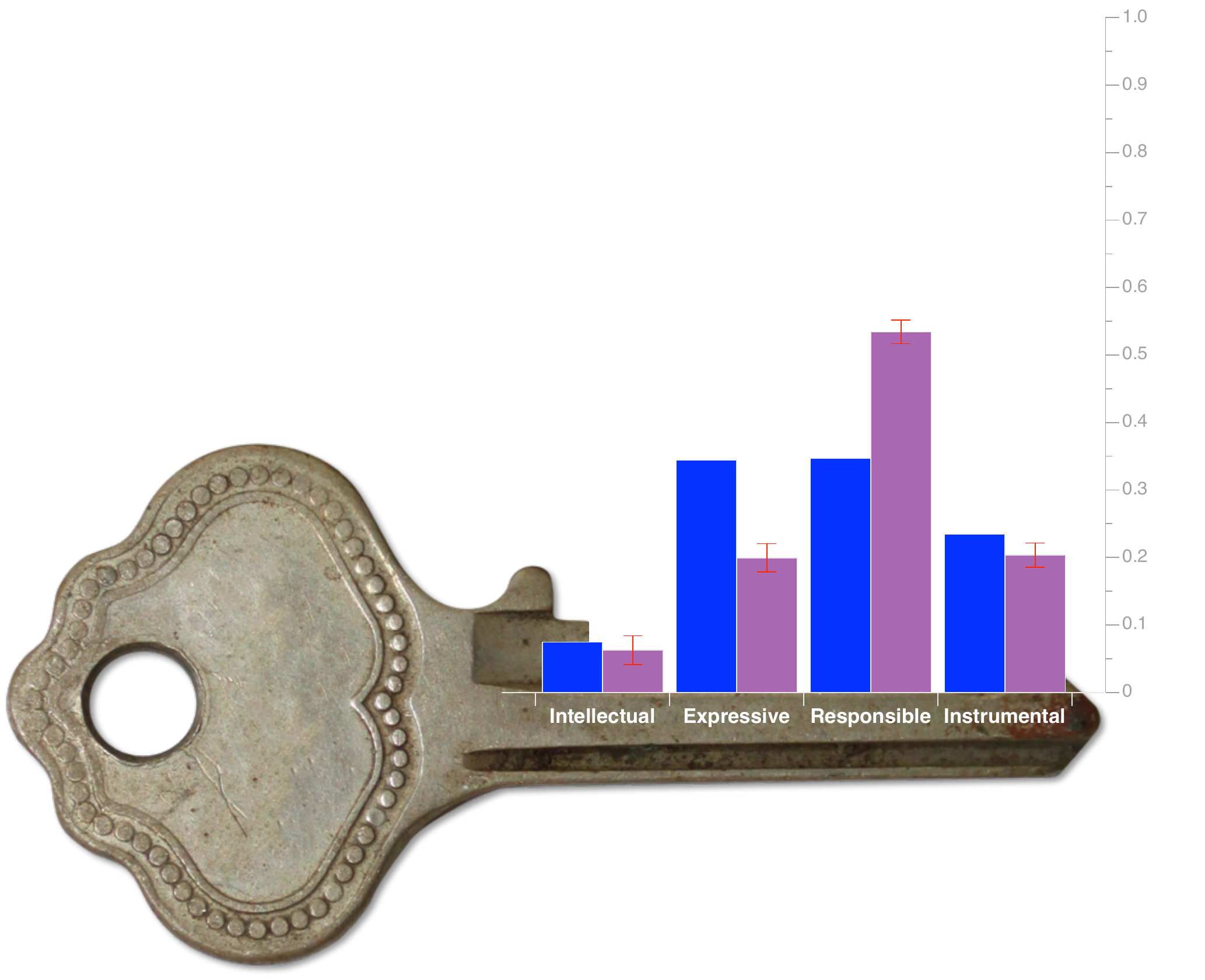}
  \caption{Keys of compatibility: Motivational Orientation (Top), Optimum of Gratification (mid) and Resulting Basic Action Types (bottom) for businessmen (left, red) and politicians (right, blue) in comparison with 3 random measures (purple). According to Tables \ref{matrix1_used} and \ref{matrix1_used_B}. E.R.=Expected Roles, I.V.=Internalized Values, R.A.=Recognition of Affections.}
  \label{fig11_b}
  \end{center}
\end{figure}

All the detected (social) actions give us the ``keys of compatibility'' associated with Motivational Orientation, Optimum of Gratification  and Resulting Basic Action for politicians and businessmen (Fig. \ref{fig11_b}).
 
\begin{table}[h]
 \caption{Conducts and effects of politicians and businessmen associated with their objects of situation (left part of the table). Capacity of politicians and businessmen for organization and to create political formula as Domination Factors. * I=Intellective, R= Responsible, E=Expressive. ** P=Power, D=Domination, R= Responsible. [Source: own elaboration].}
 \begin{tiny}
 \begin{tabularx}{250pt}{XXX|XX}
\hline 
\multicolumn{3}{c|}{ENTITY} & \multicolumn{2}{c}{DOMINATION FACTORS} \tabularnewline
\multicolumn{3}{c|}{POLITICIAN/BUSINESSMEN} & \multicolumn{2}{c}{POLITICIAN/BUSINESSMEN} 
\tabularnewline
\hline 
Need Dispositions & Conduct* & Effect** & Capacity of organization & Capacity to create political formula \tabularnewline
\hline 
Roles expectatives & I / I & P / P & high / high & high / low  \tabularnewline

Internalized values & R / R & D / R & low / low & high / low\\
 
Recognition of affect& E / E & D / D & low / high & high / low \\\hline

\end{tabularx}
\end{tiny}
\label{matrix_empre}
\label{pvsb}
\end{table}

A clear difference  between  groups can be observed. In businessmen (Fig. \ref{fig11_b}, top row, left) relations depends on Expected Roles, while those that depend on Internalized Values and Recognition of Affect are less frequent. This result  is quite different than the observed behavior between politicians (Fig. \ref{fig11_b}, top row, right) where Internalized Values seems to dominate the relations inside the group. With respect to the expected  gratification return, businessmen  and politicians  in the nucleus of power seem to act  in search of domination and  power; however, in the  random  samples, politicians  seem more likely  to look for power.  Finally,  with respect  to the  types of Resulting Basic Action,  the analysis shows that businessmen  (bottom row, left) have Instrumental behaviors,  followed by Expressive  and Intellectual. On the other  hand,  the behaviors  of politicians  (bottom row,  right) are  preferentially Expressive  or Responsible.  In most of the cases, random  samples show similar behaviors respective to their corresponding real data,  denoting again, robust  results.

\begin{figure}[htp]
\begin{center}
  \includegraphics[width=2in]{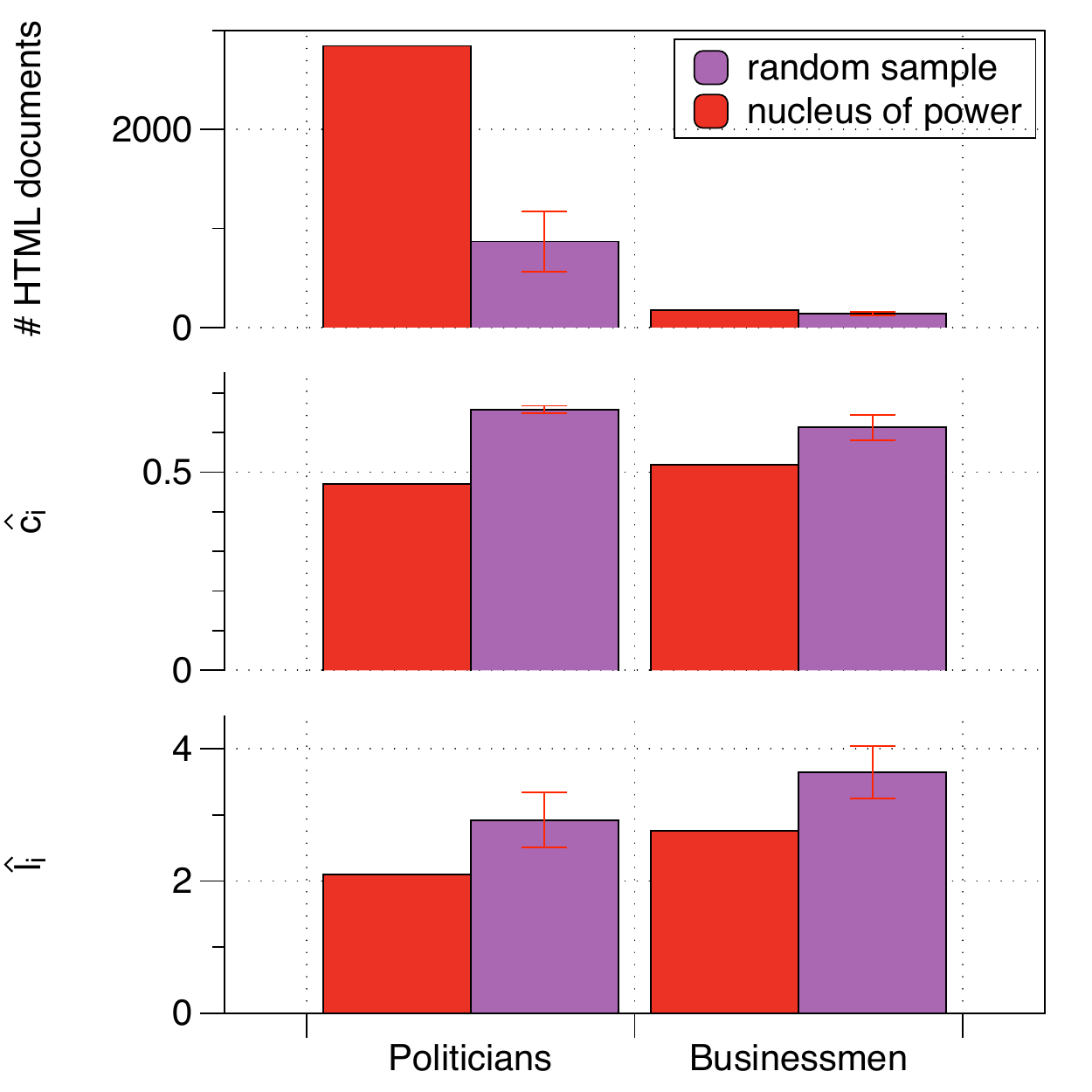}
  \caption{Politicians \textit{vs} businessmen. Average shortest path length $\hat{l}_i$ from  entity $i$ to the rest (bottom), average clustering coefficient of entity $i$ (mid) and average number of HTML documents of entity $i$ ``saying something'' (top), in comparison with 3 random samples (purple bars).}\label{fig12}
  \end{center}
\end{figure}

Although the keys herein presented already provide valuable insight, it is possible to go deeper in these  results because, according  to  sociological theory,  there  are conducts  and  effects  (results) associated  to \textit{Objects of Situation} (Table \ref{pvsb}, left column). For the  two most  representatives motivational orientations of both  groups,  Expected Roles  and  Internalized Values (see Fig. \ref{fig11_b}, top row), there are \textit{Intellective} and \textit{Responsible} conducts associated \cite{Sanchez}. \textit{Power} is the effect associated with the \textit{Expected Roles} for both groups, however, a key difference appears  in the  effect of  \textit{Internalized Values}: for politicians, it is the \textit{Domination}, but for businessmen, it is \textit{Responsibility}.

Furthermore, there are \textit{Domination Factors} \cite{Mosca,Mosca2} (Table \ref{pvsb}, right column) such as ``capacity of organization'' and ``capacity to create political formula'' associated to conducts and object. The theory summarized in the  table supposes that the difference between  politicians  and businessmen  is in their  ability  to create  political formula.   In order  to test this,  we used the generated topological  entity information as an approximation of their  capacity  of organization.  Thus,  we say that this  capacity is related  with the ``size'' of their  ``world''. If their ``world'' (nearby  people) is small, we say the people have a high capacity  of organization. According to this definition we use two  classic metrics  to gauge this  ``world size'': the  mean  cohesiveness  of their neighborhood  ($\hat{c_i}$), and the average length of the shortest connections between them  ($\hat{l_i}$). As the bottom and  middle  plots in Figure \ref{fig12} show, both  groups  have  relatively the  ``same capacity''.

\begin{figure}[htp]
\begin{center}
  \includegraphics[width=3in]{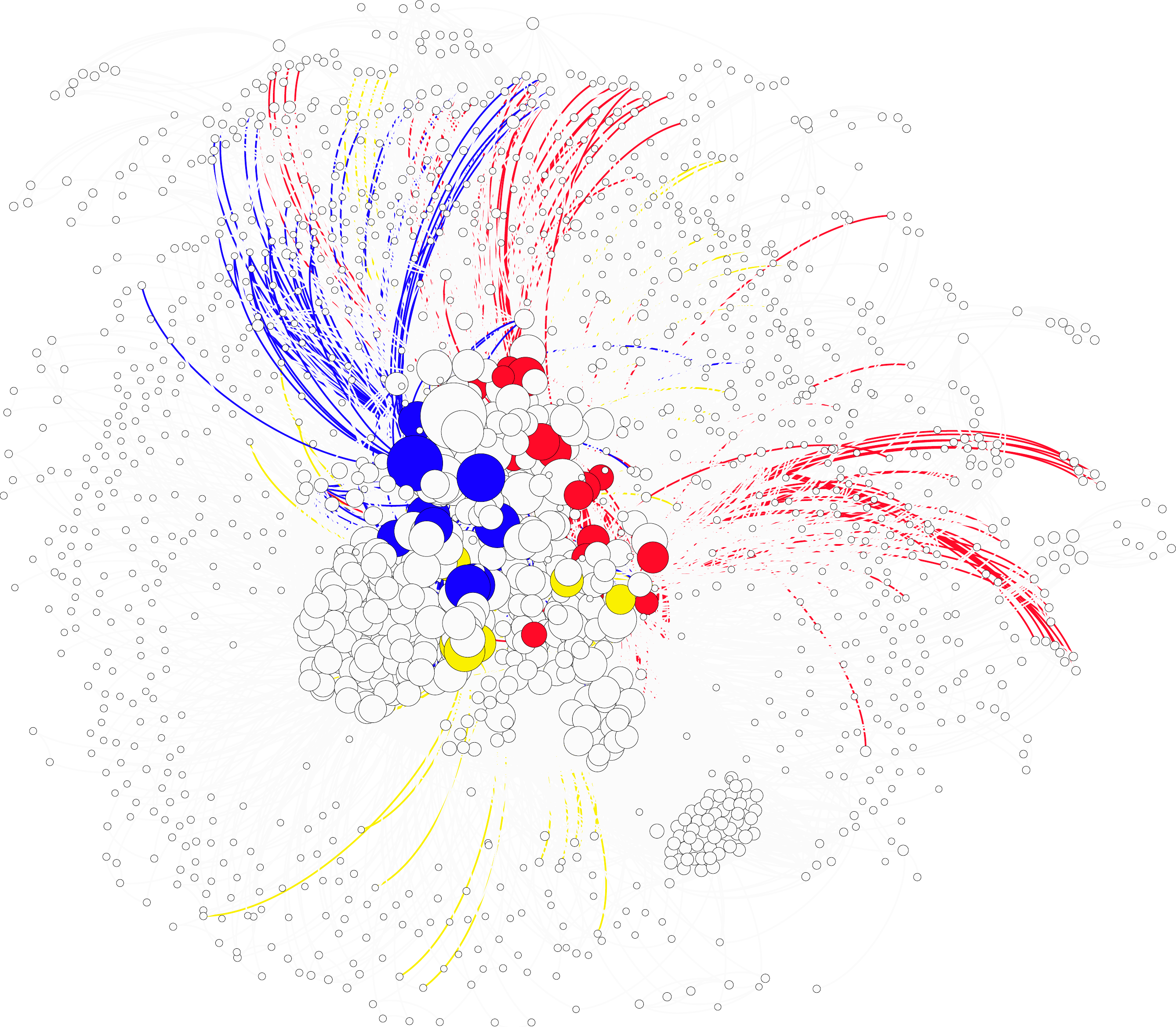}
  \caption{Three major communities of Chilean power network. Layout shows the powerful in the center. The nucleus of power (40 people) is colored according the profile of the person: businessmen (red), liberal politician (yellow) and conservative politician (blue). }
  \label{fig13}
  \end{center}
\end{figure}

Since these results were so similar, another approach was needed to observe differences in the  capacity  to  create  political formula.   In  order  to  measure this difference,  the  number  of HTML  documents  that contains  words such as ``say'', ``explain'', among others,  associated  to the person were used as indicator, and  as a primary  and  very basic attempt to approach the problem.  We worked under the  assumption that phrases  such  as ``He said...''   or ``She explained...''  denotes construction of language that can lead to ``essay building'' which may mean to reach a \textit{consensus}  \cite{KLT,Rheingold} in the population or to build legitimacy \cite{Weber}. As Figure \ref{fig12} (top plot) shows, politicians are much more prolific than businessmen in this respect.

Finally,  we connected  the  profile of persons  with   social power  $P$ defined in the previous section. In order to obtain clear results, we divided the politician group in two: liberals and conservatives. Figure \ref{fig13} shows the positions of top businessmen (red) and top politicians (blue=conservatives, yellow=liberals) in the three communities described in Section \ref{chile_net}. The layout of the figure shows that the less powerful people ``orbit'' the core of power. As can be seen, a high proportion of links of conservatives politicians and businessmen go to the periphery, but  most of the links of liberals seem to be inside this core.

\begin{figure}[h]
\begin{center}
   \includegraphics[width=3in]{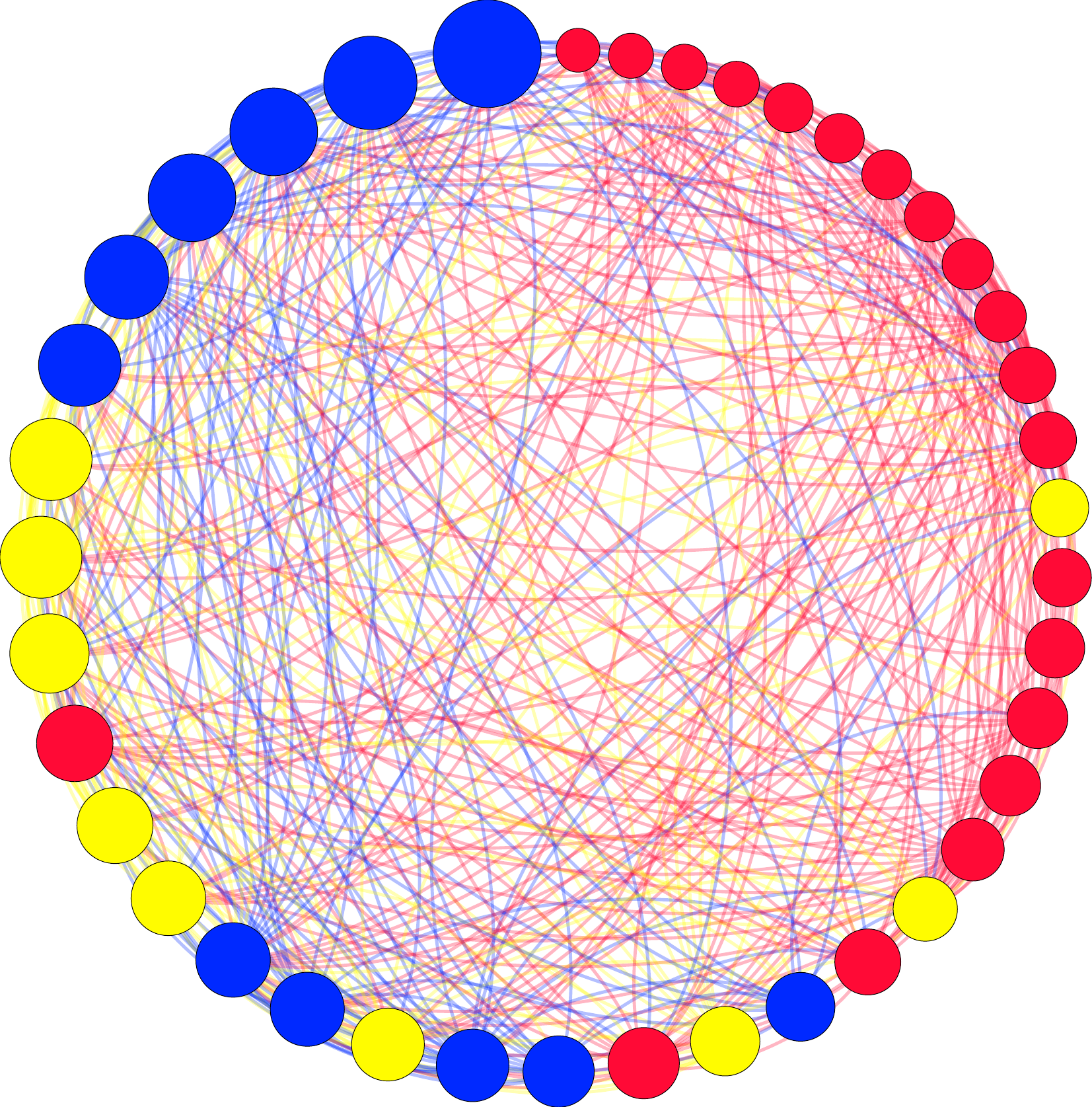}
  \caption{
  Nucleus of power isolated and sorted (in the clockwise direction) according to their $P$ power. Color nodes as in Fig. \ref{fig13}. }
  \label{fig16}
  \end{center}
\end{figure}

Once these people were isolated  from the core and sorted,  clockwise, according  to their $P$ power (Fig. \ref{fig16}), we can observe that Chilean businessmen  are clearly less powerful than politicians, especially conservatives.

\section{Discussion and conclussions}

The  results  of the  previous  sections  open the discussion about  the validity  of the original claim of sociology, that of explaining social facts, which assumes a  misleading  perception that it  is possible to  differentiate the  dimensions of subject  and  object  and  thus  to accentuate the  ``objectivity''  in the  same way as natural sciences do.  That's  what  Durkheim \cite{Durkheim,Maldonado,Alvarez,Navarro} and  other  founders  of the  Social Sciences believed. However, to  continue  to  support this  separation is difficult because  those things that we objectify  (social facts)  end up being social constructs inseparable from their attributed meanings  (interpretive values), which in turn allows for (undue) interference stemming from subjectivity (\textit{i.e.},  the  subject).   The epistemological  consequences  are varied,  and  in part,  we try  to mediate these phenomena in this study.  That said,  and through this  novel epistemological  approach to the phenomenon of power, here are some discussions and conclusions about  the results.

The social network of powerful Chilean people shows non-trivial and non-intuitive statistical properties, commonly observed in complex systems. The most remarkable of these have to do with  the  unequal distribution of power and  its links in the population, and  that these distributions  are of \textit{scale-free} character. It is also surprising to see that power in Chile is highly cohesive, as denoted by the  high  degree  of transitivity of connections.  On average,  more than half of the connections  of people in the network are also interconnected. Moreover, the power seems to be ``tight-knit'',  a real small-world where at most only a few people ($\sim$ 4) are intermediaries between any randomly  chosen pair of individuals  in the network.

The  presence  of these  properties  is very  similar  to  those  observed  in complex systems of different natures. For this reason, their presence can be explained,  at least intuitively, as an effect of their  growth and adaptation. The time and the mechanisms associated  with these  processes most likely play an important role in their emergence. The ubiquity (universality) of these  complex  properties   suggests  that there  may  exist universal  principles  underlying  the evolution  of systems,  irrespective  of their  origin.

With respect  to  this  underlying  principle  that may  generate  this  kind  of system structure, our  results  suggest  that a simple  compatibility mechanism  between people in a growing network  may engender  a social structure such as the one studied, \textit{i.e.}, strongly  inhomogeneous  in its  connectivity and  power distribution. According to Taleb \cite{Taleb}, this inequality is unexpected for most of us and would be the reason why it seems to be a non-intuitive behavior.  

According to the CAM, the compatibility of personal characteristics (see Figs. \ref{fig10} and \ref{fig14}) seem to be the ``key'' for people to belong to a elite group. But,  this mechanism  is not exclusive to the most powerful group;  it operates  at all scales, as indicated by the observation of scaling in the distribution of power  $P$  or connectivity.  As a result, there  is no single ``nucleus of power''; power concentration exists at  all levels, and even the  inequality between  powerful people  across  their  distribution is the  same. This coincides with  Dahl's  idea  of Polyarchy \cite{Dahl}, a government of many  that is structured around  a system of \textit{poliar}  balance.  As societies  develop,  they  become  more  concrete;  this concretion, then, manifests itself politically as  an  increase  in the  number  of institutions and  leaders  (the principle of ``dynamic density'' of Durkheim \cite{Ritzer}). This  leads to a situation of many overlapping  institutions that constantly need to negotiate  in order to achieve their  purposes.   To resolve any  issue that arises, powerful institutions with their own interests and different leaders have to be in agreement.


The concentration of social power in a few seems to be an emergent property of a dynamical  system.  This is not the product of a conspiracy, but the result of unregulated growth  in the  social system.   However,  according  to the  model, the  distribution of personal  characteristics plays  an  important role.   In  fact,  as \cite{CAM} exposes and under assumptions of compatibility, inhomogeneous  distributions of personal  characteristics could lead to this type of network, while homogeneous  distributions oppose concentration. This suggest that more homogeneous populations could affect, in some way, this risky\footnote{We say that concentration is risky because there are possible scenarios when the action of (or over) extremely powerful people may affect many people (many directed and undirected connections depends on them).  However, these scenarios are less probable than the ones where the action of (or over) less-powerful people.} concentration of power.  This is an important argument in favor of social reforms that ensure equality  and social inclusion.


Another interesting aspects of the proposed model is that according to Eq. \ref{eq1}, the compatibility should be more accurate as the population grows. Belonging to the  elite in large populations requires sharing  more and  more things. The research  of J.  Barbier  \cite{Barbier} on Chilean  elites between  1755 and  1800 showed that the elite was composed by approximately 347 people, representing 0.04\% of the population at that moment. Moreover, the study argues that during the last quarter of the  17$^{th}$  century  and  the  first half of the  18$^{th}$  century, the  structure of the  Chilean  elite was characterized by strong  family ties (from marriages  among the elite)  linked  to  the  possession  of titles  of nobility.    This  condition  favored the  strong  link between  power and  political  control  and  associated  business,  an issue that would continue  in the  second  half  of the  18$^{th}$  century, even though  this  same elite gave greater  importance to matrimonial relationships with  those managers  and government administrators through whom enough influence was generated to favor power and political  control.

Today,  with  a Chilean  population over 16 times  greater, belonging to  a family  of the  elite is not  enough.  Although  our results  suggest  that family relations  are important as direct  links (see Fig. \ref{fig8}), the  results  also show that not  every person that descended  from the  elite has secured  power, as was most common  in the  18$^{th}$ century.  There are many examples of people that belong to the elite without an elitist past.   Nowadays, belonging to a powerful group  means  to share  common  interests  (Expected Roles, Positioning, Functions) and  values (see Fig. \ref{fig10}). These commonalities may (or may be not\footnote{According to W. Pareto \cite{Alonso1,Marmuc,WPareto} (non-rational) psychological residues can be involved in this distinction. Pareto understand that social  interaction is directly linked  to  psychological predispo- sitions.  People are  linked  not  by  ``rational action''   or  ``instinctive actions'', but by  ``non-rational actions''. The  latter are,  in his opinion, what normally occurs. Such  actions are  covered by a veil that gives  appearance of logic (which Pareto called  ``derivation''), but if that veil is removed, only ``residues''  remain (psychological predispositions).  Pareto distinguishes different  types  of residues that remain in the  elite,  highlighting ``residues  of persistence'' that  causes those who  have  it to  prefer change over  tradition.  These people  also  have  no religious attachment.  On  the  other hand, people  that have  ``residues  of combination'', tend to  preserve a  certain state of things, to value  the  traditions and  have  a certain religious attachment. Using  a metaphor of Machiavelli, the first  may  be identified as ``lions'' (conservatives), while  the  latter as ``foxes'' (liberals).}) have  been  inculcated during  (similar) formations  in their  homes or in a few groups  of schools and  universities  (see results of Fig. \ref{fig8}). Therefore, people  could be linked in various  nuclei  of power due to the  interpretation of senses and meanings  attached to the mutual construction of symbols, which contain  different ``senses'' linked to \textit{objects of gratification} of power and domination, associated  with preferably  expressive  and  instrumental conducts  (see Fig. \ref{fig14}).  The results showing that powerful people link to persons with similar power (see Section \ref{model}) dramatically reflect this process, although, as a remarkable outcome of our analysis, this phenomenon  was shown to occur between  less powerful people as well.


The  effects of relation  by likeness and similarity, accompanied by the imitation phenomenon \cite{Mosca,Mosca2}, are  the  key to  understanding the phenomenon  of attraction. Although this was illustrated by W. Mills in his study  of  ``Power Elite'' in the United States \cite{Mills},   it  had  already  been  presented as a caveat  by R.  Michels in his ``law of the psychological metamorphosis of the leader''. This law reflects the inevitability  that powerful people  end up acting  and  behaving according  to  the  parameters of the  elite  and,  indeed,  thinking  similarly and exhibiting an attitude of conserving positions of power.  This is one possible interpretation from looking at the nucleus (nuclei)  of power.

When we study  the two groups of powerful people\footnote{Our suspicion is that, since its inception, the power in Chile has focused in the strong link between politics and business, which has led us to prioritize our research on these two categories.}, businessmen and politicians, more profoundly,  the results show that their  ``keys of compatibility''  are different (see Fig. \ref{fig11_b}). This is not at all surprising , since power has a different meaning for each group, despite the fact that in both cases we are talking about a power that emerges from relations, a \textit{social capital} \cite{Adler}.  Economic power is projected  through trade  relations  that allow actors  access to goods, which has an \textit{instrumental} character to their  purposes \cite{Nahapiet, Rodrigo}. On the other hand,  political power does not refer to possession of goods or instruments, but  the  ability  to control the  system  of relations.   Political  power is the  mobilization  of all the  relational context  as a means.   Indeed,  our results  show that powerful businessmen,  mostly  engineers  or economists,  seem to be dominated by a compatibility based on Expected Roles (interest), while politicians, mostly lawyers, seem to be moved by compatibility of Internalized Values (see Fig. \ref{fig11_b}). These results would again be explained  by sociological theories of social power.  According to M. Weber \cite{Weber}, power is the imposition  of one's will over another, even against  all resistance,  and  ultimately rests  on the  use of force or threat thereof. However, it is much smarter to ensure prompt and sincere obedience, which Weber himself appointed ``legitimate domination'': The power required to legitimize aims to capture the belief of the people, which is where social legitimacy lies. Hence, politics become a more efficient instrument to build legitimacy, ``to convince''. In that sense, there  is a greater  ability  of politicians  to undertake this task  than  businessmen (see Fig. \ref{fig12}, upper plot). Unlike the nature of politics, the economy  seems to  rest  on a principle  of rationality linked  to  obtaining optimal  profits that guide social actions  (supply  and  demand). In this latter logic, legitimacy  building   does not seem a requirement \textit{sine qua non}, and thus reduces to some extent efforts to hold onto the ``great narratives''  that contain  principles  and rating  of ``political economy''.  This  difference in constructing legitimacy sheds some light differences observed between  politicians  and businessmen  in our work.

Furthermore, businessmen generally seem to follow a sectoral logic. Their interactions are fundamentally cognitive  and  are related  to expected roles, functions,  and  positions. In that sense, the  search for power seems to be more explicit.  Politicians, for their part,  seem to generally  obey a sectoral  logic associated  with obtaining control  over processes,  planning,  and  distribution within the  social system.   In their  search  for social legitimacy,  their interactions seem to be related  to domination, which is why their  behavior  seems to follow value orientations of social and moral responsibility.  However, in both  groups, affect orientations operate  transversely  and equivalently. Usually, however, actors do not usually make this kind of orientation ``expressive''. This is one possible explanation of our analysis, in which the affect orientation appears to play a less important role (see Figs. \ref{fig10} and \ref{fig11_b}).

In light of the  results  and  discussions above,  there  are some ideas that want further development. One of these is related to the Kornhauser's categories \cite{Korn}: ``accessibility of elites'' (how permeable  the elites are to the influence of non-elites) and ``availability of non-elites'' (how non-elites are mobilized by elites). The role played by Chilean society (the non-elite) since 2011 \cite{Laszlo,Vidal1,May}, seems to be the result  of a process that has transformed Chilean  society into a more pluralistic one. In such societies,  the  accessibility  of elites tends to be increasingly  high (high ability  to  be permeated by non-elite)  and  the  ability  of non-elite  to be mobilized (by elites) is increasingly  difficult.  This makes us notice that linkages of power also tend  to become part  of this dynamic.

Finally, and maybe the most remarkable underlying conclusion of our work, is that power seems to be in the network, in the connections, in the processes, in the codes, in the communication, in everything; the concentration (at  different scales) that we presented is nothing more than the ``instrumental'' organization of power, its efficient administration with respects to purposes related  to control and maintaining the  network  itself.  A view in agreement with the observation made recently by Alvaredo \textit{et al.} \cite{Alvaredo} respect to the economic elites. This  represents a different approach to  social power  in comparison  with  the  classic view related  to  obtaining certain  power that only a few are privy to (``Theory of Elites''). Power seems to be rather a phenomenon of complex social systems, part of its substantive properties, not just  the result  of petty  intentions of mastery of the ``few over the many'', even when these seem to be the ``real'' expression that we commonly observe and experience. This opens an interesting next  research  that we want to develop in the future.


\section*{Acknowledgments}

This work was financed by projects: ``Dise\~no de Redes'' from INRIA-Chile and Universidad T\'ecnica Federico Santa Mar\'ia, and CONICYT SOC-1101, ``Anillo en Complejidad Social'' of Universidad del Desarrollo de Chile. We also want to thank Poderopedia.org for the data provided, and we appreciate the comments and suggestions made by Professors Arturo Fuenzalida, Mason Taylor and Miguel Paz, which contributed significantly to better expose the results of our research.  



\begin{thebibliography}{99}

\bibitem {Adler}
\newblock P.S. Adler and S.W. Kwon, 
\newblock  Social Capital: Prospects for a new concept,
\newblock \emph{The Academy of Management Review}, \textbf{27} (2002), 17--40.

\bibitem {Alonso1}
 J. Alonso,
  \emph{Pareto},
 1$^{st} edition$, Edicol, M\'exico, 1977. 


\bibitem{Python} 
      J. Alstott, E. Bullmore and D. Plenz,
      Powerlaw: a python package for analysis of heavy-tailed distributions,
      \textit{arXiv:1305.0215v3}.

\bibitem {Alvaredo}
 F. Alvaredo, A. B. Atkinson, T. Piketty and E. Saez,
  The top 1 percent in international and historical perspective,
 \emph{Journal of Economic Perspectives}, \textbf{27} (2013), 3--20.

\bibitem {Alvarez}
 G. \'Alvarez,
  Caos/Complejidad, Fractales e Identidades sociales,
 \emph{Raz\'on y Palabra}, \textbf{79} (2012). Available from: \url{http://www.redalyc.org/articulo.oa?id=199524411055}.
     
\bibitem {Aron} 
 R. Aron,
  \emph{La sociolog\'ia alemana contempor\'anea},
  Paid\'os, Buenos Aires, 1965. 

\bibitem {Ball}
\newblock P. Ball,
\newblock  \emph{Critical Mass},
\newblock Farrar, Straus and Giroux, New York, 2004. 

\bibitem{BA} 
      A-L. Barab\'asi and R. Albert,
      Emergence of scaling in random networks,
      \emph{Science}, \textbf{286} (1999), 509--512.

\bibitem {Bertalanffy}
 L. von Bertalanffy,
  \emph{Teor\'ia General de los Sistemas},
 M\'exico, (1991). Available from:
\url{http://es.scribd.com/doc/135617671/TGS--Bertalanffy--pdf?}

\bibitem{Barbier} 
      J. A. Barbier,
      Elite and cadres in bourbon Chile,
      \emph{The Hispanic American Historical Review}, \textbf{52} (1972), 416--435.

\bibitem {Bartolo} 
 J. Bascompte and B. Luque,
  \emph{Evoluci\'on y Complejidad},
 Publicacions de la Universitat de Val\'encia, Val\'encia, 2012. 

\bibitem{Bascompte2} 
      J. Bascompte and P. Jordano,
      Plant-animal mutualistic networks: the architecture of biodiversity,
      \emph{Anual Review of Ecology, Evolution and Systematics}, \textbf{38} (2007), 567--593.

\bibitem {Bendix}
 R. Bendix,
  \emph{Max Weber and intellectual portrait},
 Doubleday \& Company, Inc., New York, 1962. 

\bibitem{Blon} 
      V.D. Blondel, J.-L. Guillaume, R. Lambiotte and E. Lefebvre,
      Fast unfolding of communities in large networks,
      \emph{J. Stat. Mech.}, \textbf{2008} (2008), P10008.

\bibitem {Buchanan}
\newblock M. Buchanan,
\newblock  \emph{Ubiquity},
\newblock Three River Press, New York, 2000. 

\bibitem{CAM} 
      J.P., C\'ardenas, M. Mouronte, J. C. Losada and R.M. Benito
      Compatibility as underlying mechanism behind the evolution
of networks,
      \emph{Physica A}, \textbf{389} (2010), 1789--1798.

\bibitem {Dahl}
 R. Dahl,
  \emph{La Poliarqu\'ia: Partcipaci\'on y Oposici\'on},
 Tecnos, 2009.

\bibitem {Dahrendorf}
 R. Dahrendorf,
   \emph{La ley y el orden},
 1$^{st}$ edition, Editorial Civitas, 1994. 

\bibitem {Deploige}
 S. Deploig\'e,
  \emph{El Conflicto de la moral y de la Sociolog\'ia},
  McGraw-Hill, M\'exico, 1920. 

\bibitem{Dor} 
      S.N. Dorogovtsev and J.F.F. Mendes,
      \emph{Evolution of networks. From biological nets to the internet and www},
      Oxford University Press, New York, 2003.

\bibitem {Durkheim}
 E. Durkheim,
  \emph{Las reglas del m\'etodo sociol\'ogico},
  Dedalo, Buenos Aires, 1964.

\bibitem {ER}
 P. Erd\H{o}s and A. R\'enyi,
 On random graphs,
  \emph{Publicationes Mathematicae (Debrecen)},  \textbf{6} (1959), 290--297.

 \bibitem {Foucault}
 M. Foucault,
  \emph{Microf\'isica del poder},
 2$^{nd}$ edition, La Piqueta, Madrid, 1980. 

\bibitem {GdC}
 O. Gonz\'alez del Cardenal,
   \emph{El Poder y la Conciencia. Rostros personales frente a poderes an\'onimos},
 2$^{nd}$ edition, Espasa Calpe, Madrid, 1985.

\bibitem{Gra} 
      M. S. Granovetter,
      The strength of weak ties,
      \emph{American Journal of Sociology}, \textbf{78} (1973), 1360--1380.

\bibitem{Har} 
      F. Harary, R. Norman and D. Cartwright,
      \emph{Structural Models},
      Wiley, New York, 1965.


\bibitem {KLT}
 P. Klimek, R. Lambiotte and S. Thurner,
  Opinion formation in laggard societies,
 \emph{Europhysics Letters}, \textbf{82} (2008), 28008.

\bibitem {Korn}
 W. Kornhauser,
  \emph{The Politics of Mass Society},
 Routledge, New York, 1959.


\bibitem {Laszlo}
 E. Laszlo,
  \emph{La Gran Bifurcaci\'on. Crisis y oportunidad; anticipaci\'on del nuevo paradigma que est\'a tomando forma},
 Gedisa, Barcelona, 1990.

\bibitem {Lenski}
 G. Lenski,
   \emph{Poder y Privilegio. Teor\'ia de la estratificaci\'on Social},
 Paid\'os, Buenos Aires, 1969.


\bibitem {Lopez}
 R. L\'opez, E. Figueroa and P. Guti\'errez,
  \emph{La parte del Le\'on: Nuevas estimaciones de la participaci\'on de los s\'uper ricos en el ingreso de Chile},
 Universidad de Chile, Departamento de Econom\'ia. Serie de Documentos de Trabajo, (2013). Available from:
\url{http://www.econ.uchile.cl/uploads/publicacion/306018fadb3ac79952bf1395a555a90a86633790.pdf}.


\bibitem {Maldonado}
 C. Maldonado,
  Complejidad y ciencias sociales. el problema de la medici\'on de los sistemas sociales humanos,
 \emph{Complejidad de las ciencias y ciencias de la complejidad}, (2006), 1--145.



\bibitem {Marmuc}
 B. Marmuc,
  Ideolog\'ia, matem\'aticas y ciencias sociales, W. Pareto y G. Sorel y la ambig\"uedad en la comparaci\'on de las desigualdades,
 \emph{Revista de Metodolog\'ia de Ciencias Sociales}, \textbf{5} (2013), 11--28.

\bibitem{May} 
      A. Mayol,
      \emph{El derrumbe del modelo},
      LOM, Santiago, 2013.

\bibitem{Mel} 
      M. Mitchell,
      \emph{Complexity. A guided tour},
      Oxford University Press, New York, 2003.

\bibitem {Michels}
 R. Michels,
  \emph{Los Partidos Pol\'iticos},
  Amorrortu, Buenos Aires, 1976. 

\bibitem {Mills}
 W. Mills,
  \emph{La elite del poder},
  9$^{th}$ edition, Fondo de Cultura Econ\'omica, M\'exico, 1988. 

\bibitem {Molina}
 J. Molina-Cano,
 Las nociones de mando y obediencia en la teor\'ia pol\'itica de Julien Freund,
  \emph{D\'ikaion}, \textbf{23} (2009). Available from: 
\url{http://dikaion.unisabana.edu.co/index.php/dikaion/article/view/1550/2126}.




\bibitem {Mosca}
 G. Mosca,
  \emph{La Clase Pol\'itica},
 2$^{nd}$ edition, Fondo de Cultura Econ\'omica, M\'exico, 1980.

\bibitem {Mosca2}
 G. Mosca,
  \emph{Elementi di Scienza Politica},
 Reproduction of the original version. Available from: 
\url{http://americo.usal.es/iberoame/sites/default/files/laclasepolitica.pdf}.

\bibitem{Nahapiet} 
     \newblock J. Nahapiet and S. Ghoshal,
     \newblock Social capital, intellectual capital, and the organization advantage,
     \newblock  \emph{The Academy of Management Review}, \textbf{23} (1998), 242--266.


\bibitem {Navarro}
 P. Navarro,
  \emph{El Fen\'omeno de la complejidad social humana},
 Curso de Doctorado Interdisciplinar en Sistemas Complejos, (1996). Available from: 
\url{http://home.dsoc.uevora.pt/~eje/complexidade_social.html}.

\bibitem{New} 
      M.E.J. Newman,
      The structure and functions of complex networks,
       \emph{SIAM Review}, \textbf{45} (2003), 167--256.

\bibitem{Newb} 
      M. E. J. Newman, A-L. Barab\'asi and D. Watts,
       \emph{The structure and dynamics of networks},
      Princeton University Press., New York, 2006.

\bibitem{Asso} 
      M.E.J. Newman,
      Assortativity mixing in networks,
      \emph{Physical Review Letters}, \textbf{89} (2002), 208701.


\bibitem{Newb} 
      M. E. J. Newman, A-L. Barab\'asi and D. Watts,
      \emph{The structure and dynamics of networks},
      Princeton University Press., New York, 2006.

\bibitem {Nocera}
 P. Nocera,
  \emph{El debate Gabriel Tarde y Emile Durkheim. De las disparidades iniciales expl\'icitas a las convergencias tard\'ias impl\'icitas},
 Available from: 
\url{http://webiigg.sociales.uba.ar/iigg/jovenes_investigadores/5jornadasjovenes/EJE9/Mesa\%20Teoria\%20Sociologica\%20Clasica\%20I/NOCERA_Pablo_d.PDF}.



\bibitem{PR} 
      L. Page, S. Brin, R. Motwani and T. Winograd,
      \emph{The PageRank citation ranking: Bringing order to the Web},
      1999. Available from: \url{http://dbpubs.stanford.edu:8090/pub/showDoc.Fulltext?lang=en&doc=1999-66&format=pdf}.


\bibitem {WPareto}
 W. Pareto,
  Rese\~na de su obra,
 \emph{Socio-Ciencia}, (2009). Available from:
\url{http://socio--ciencia.blogspot.com/2009/10/wlfredo--pareto--una--resena--de--su--obra.html}.

\bibitem {chilepoder}
 PNUD-Chile,
  \emph{La \'elite chilena y la dif\'icil conducci\'on del desarrollo},
 PNUD, 2004. Available from:
\url{http://www.centrodesarrollohumano.org/pmb/opac_css/index.php?lvl=notice_display\&id=923#.U3DOTtyodrQ}.

\bibitem {Parsons}
 T. Parsons and E. Shils,
  \emph{Toward a General Theory of action},
 Harper \& Row, publishers, New York, 1962.

%


\bibitem {Rheingold}
 H. Rheingold,
 \emph{Multitudes Inteligentes},
 Gedisa, Barcelona, 2004.

\bibitem {Ritzera}
 G. Ritzer,
  \emph{Teor\'ia Sociol\'ogica Cl\'asica},
 1$^{st}$ edition, Imprenta San Francisco de Sales,  Madrid, 1920.

\bibitem {Ritzerb}
 G. Ritzer,
  \emph{Teor\'ia Sociol\'ogica Contempor\'anea},
 McGraw-Hill, M\'exico, 1994.


\bibitem {Ritzer}
 G. Ritzer,
  \emph{Teor\'ia Sociol\'ogica Cl\'asica},
 McGraw-Hill, M\'exico, 1994.




\bibitem {Rodrigo}
 \newblock P. Rodrigo and D. Arenas,
     \newblock La nueva gobernanza pol\'itica y las colaboraciones intersectoriales para el desarrollo sostenible,
     \newblock \emph{Innovar}, \textbf{24} (2014), 197--210.


\bibitem {Runciman}
 G. Runciman,
  \emph{Cr\'itica de la filosof\'ia de las ciencias sociales de Max Weber},
 Fondo de Cultura Econ\'omica, M\'exico, 1976.


\bibitem {Sanchez}
 F. Sanchez L\'opez,
  \emph{Sociolog\'ia de la Acci\'on. Introducci\'on a la obra de Talcott Parsons},
 Instituto Balmes de Sociolog\'ia, Consejo Superior de Investigaciones Cient\'ificas, Madrid, 1964.


\bibitem{Sole}
      R.  Sole,
      \emph{Redes complejas. Del genoma a Internet},
      Tusquets Eds., Barcelona, 2009.


\bibitem {Thom}
 R. Thom,
  \emph{Esbozo de una semiof\'isica. F\'isica aristot\'elica y teor\'ia de las cat\'astrofes},
  Gedisa, Barcelona, 1990.

\bibitem {Taleb}
 N. N. Taleb,
  \emph{The Black Swan},
 Random House, New York, 2010.

\bibitem {Touchard}
 J. Touchard,
  \emph{Historia de las Ideas Pol\'iticas},
 Tecnos, Madrid, 1987. 


\bibitem{Vidal1}
     \newblock G.  Vidal,
     \newblock \emph{Chile Ante una Encrucijada?},
     \newblock 2011. Available from: \url{http://www.ugm.cl/info-alumnos-y-alumnas/chile-ante-una-encrucijada/}.
    
\bibitem {WS}
  D.J. Watts and S.H. Strogatz,
      Collective dynamics of small-world networks,
      \emph{Nature}, \textbf{393} (1998), 440--442.

\bibitem {Weber}
 M. Weber,
  \emph{Econom\'ia y Sociedad},
 Fondo de Cultura Econ\'omica, M\'exico, 1964.


\end{thebibliography}
\end{document}